\documentclass[10pt]{article}
\usepackage[utf8]{inputenc}
\usepackage[margin=1.0in]{geometry}
\usepackage[english]{babel}
\usepackage{bibentry}
\usepackage{natbib}
\setcitestyle{numbers,open={[},close={]}}
\usepackage{array}
\usepackage{hyperref}
\hypersetup{
    colorlinks=false,
    linkcolor=blue,
    filecolor=magenta,      
    urlcolor=black,
    citecolor=blue
}
\usepackage{graphicx,kantlipsum,setspace}
\usepackage{lipsum}
\usepackage{ragged2e}
\usepackage{cleveref}
\usepackage[dvipsnames]{xcolor}
\usepackage{tikz}
\usepackage{pgfplots}
\pgfplotsset{compat=1.14}
\usepackage{pgfplotstable}
\usepackage{caption}
\usepackage{subcaption}
\usetikzlibrary{calc}
\title{Energy Correction in Reduced SiD Electromagnetic Calorimeter}
\author{Lucas Braun, Jason T. Barkeloo, James E. Brau and Christopher T. Potter\\\\University of Oregon, Eugene}
\date{\vspace*{-1cm}}

\begin{document}
%\begin{flushright}
%February 13, 2020
%\end{flushright}

\maketitle

\begin{tikzpicture}[remember picture,overlay]
          \node[anchor=north east,inner sep=0pt] at ($(current page.north east)-(3.cm,1.5cm)$) {
       {\large \today}
        };
\end{tikzpicture}

\begin{abstract}
    SiD is a robust, silicon-based detector proposed for the International Linear Collider (ILC). SiD employs a sampling silicon-tungsten electromagnetic calorimeter (ECal) to accurately measure the energies of electrons, positrons, and photons produced in collisions, and to contribute to jet energy measurements through the particle flow technique. Due to the nature of the detector and its design constraints, a portion of the electron, positron, and photon energy exits the ECal undetected. Here, we establish a methodology for estimating the exiting energy and correctly determining the energies of electrons (and photons) in the ECal by analyzing patterns in the total energy deposition in each layer using neural networks. We studied a reduced calorimeter design with fewer layers (16 thin layers, 8 thick layers) than the proposed SiD design (20 thin layers, 10 thick layers) to evaluate if the resulting energy leakage could be determined and corrected. We evaluated the effectiveness of the correction on various electron energies and incidence angles in a Geant4 simple stack as well as in a modified version of the full DD4hep SiD model. The correction methodology showed significant improvement in measurement accuracy and resolution over uncorrected events, especially at high energies and shallow angles. These results provide a basis for effective energy correction in the SiD ECal and suggest that precise energy measurements with a smaller, less expensive ECal are viable.
\end{abstract}

\tableofcontents

\section{Introduction}

\subsection{SiD Detector}
SiD is a robust, silicon-based detector proposed for precise measurements at the International Linear Collider ILC \cite{behnke_international_2013}. Silicon-based tracking and calorimetry, combined with a 5T superconducting solenoid, allow for effective particle flow algorithm (PFA) implementation. This yields strong jet energy resolution, allowing for effective analysis of Higgs decay. For PFA implementation, SiD calorimeters must have sufficient depth to avoid significant leakage, thereby ensuring energy measurement accuracy.

\subsection{SiD Electromagnetic Calorimetry}
\begin{figure}
    \centering
    \includegraphics[scale=0.15]{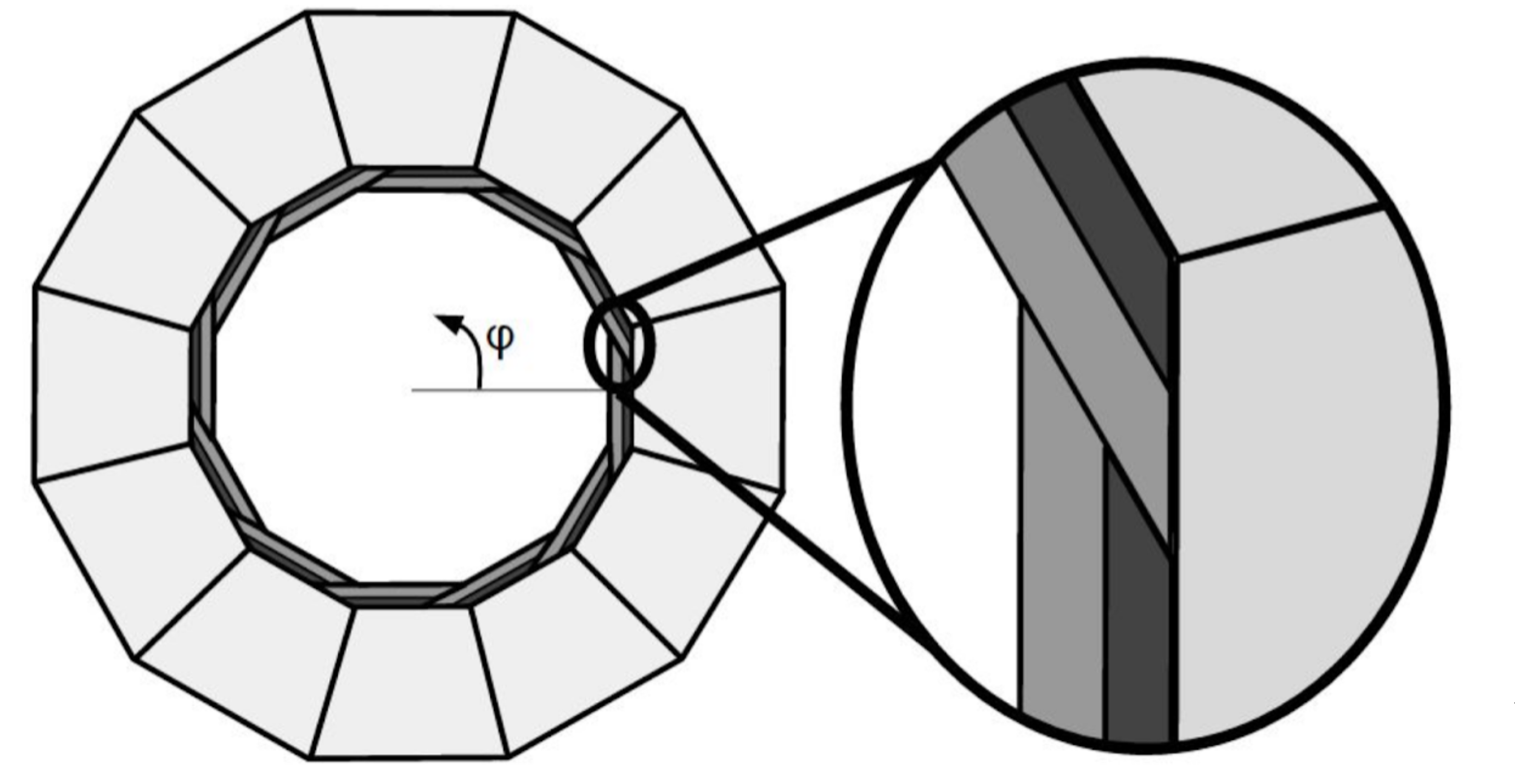}
    \caption{Diagram of SiD ECal surrounded by HCal. The medium and dark gray show regions containing thin and thick ECal layers, respectively. The full SiD model implemented here uses this geometry, most notably the overlapping regions which occur for certain $\phi$ values.}
    \label{fig:ECalOverlap}
\end{figure}

The SiD electromagnetic calorimeter (ECal) is composed of alternating tungsten alloy and silicon layers. The silicon layers detect energy deposits while the tungsten advances the electromagnetic shower, absorbs energy without detecting, and is of variable thickness. The nominal SiD ECal \cite{behnke_international_2013} calls for 20 thin (2.5mm) and 10 thick (5.0mm) tungsten layers, 30 silicon layers, and a 1.25mm readout gap, for a total depth of 26$X_0$. This design has an electron and photon energy resolution of $\Delta E/E = 0.17/\sqrt{E} \oplus 0.01$. The ECal barrel uses twelve modules for ease of construction \cite{behnke_international_2013}. The modules overlap, preventing gaps and providing structural support (see figure \ref{fig:ECalOverlap}). This creates regions where thin layers extend to the outside of the calorimeter.

Optimized cost estimates for the SiD ECal \cite{behnke_international_2013} place the ECal as one of the most expensive SiD components. High ECal materials cost motivates using fewer layers. However, this would worsen the existing problem of energy leakage from the ECal into the HCal for high-energy and late-showering electrons.

\subsection{SiD ECal Energy Leakage Correction}
Here we developed a methodology for improving the accuracy of the measurements of electron energies in a reduced ECal by analyzing the total energy deposited in each silicon layer to account for leakage. We analyzed the performance of first-order corrections that used particle shower properties which correlated with energy loss in an energy-independent way. We then utilized neural networks to improve the correction by considering the total energy deposited in each ECal layer.

Our model utilized a simple-stack ECal design simulated in Geant4 \cite{allison_geant4_2006,allison_recent_2016}, as well as a modified version of the full DD4hep-described SiD model \cite{frank_dd4hep:_2014}. The simple stack consisted of alternating 2.24 mm thick pure tungsten layers and 0.32 mm thick silicon pixel layers with no gaps in between. The stack contained a total of 60 silicon layers and 60 tungsten layers. Similarly, the modified full SiD model was extended to 60 thin silicon and tungsten layers while preserving the orginial geometric configuration (see figure \ref{fig:ECalOverlap}). Though the SiD TDR baseline design calls for 20 thin (2.5 mm) tungsten alloy layers followed by 10 thick (5.0 mm) tungsten alloy layers, including 60 thin layers allowed for shower development beyond the baseline design to be analyzed \cite{behnke_international_2013}.

To emulate different ECal designs, different sections of the energy deposits were considered. A nearly-ideal ECal design consisting of all 60 thin layers (the 60 thin or 60+0 design) was used to approximate the theoretically ideal electron energy deposit. We simulated the current SiD design (the nominal or 20+10 design) by only using data from simulated layers 1 through 20 and even-numbered layers 22 through 40. A smaller design using only 16 thin and 8 thick layers (the reduced or 16+8 design) was simulated using data from layers 1 through 16 and even-numbered layers 18-32. Deposits in the odd-numbered layers in the ECal were approximated by fitting the parameterized formula $f(t)=Ct^{\alpha}e^{-\beta t}$. Our research focuses on providing energy correction to the 16+8 design to better assess the correction methodology's capabilities, as well as to evaluate the potential to reduce the size of the calorimeter and cut costs.

\section{Results}
Here we present the performances of first-order and neural network (NN) energy correction methodologies under different conditions. The first-order correction consisted of finding particle shower properties which correlated with energy loss in an energy-independent way and using them to correct electron event energies. The first-order correction was tested on data sets containing events with constant electron energies fired normal to the ECal. The neural-network-based energy correction methodology was tested on a variety of single-electron data sets. Data sets for both training and testing were divided into four categories:

\begin{enumerate}
    \item Electrons fired into the ECal at a constant energy and constant angle of incidence,
    \item Electrons fired at a random distribution of energies between either 20 and 300 GeV or 10 and 300 GeV but at a constant angle of incidence,
    \item Electrons fired at a constant energy but at a random distribution of angles between 0$^{\circ}$ and 50$^{\circ}$, and
    \item Electrons fired at a random distribution of energies between 20 and 300 GeV and at a random distribution of angles between 0$^{\circ}$ and 50$^{\circ}$.
\end{enumerate}

In cases where constant electron energies were used to both train and evaluate the model, electrons fired with the same energy which was being evaluated were excluded from the training data to avoid biasing the results.

Note: unless otherwise stated, the error for resolution values shown here is given by:
\begin{equation}
    \sigma_{E}/(E_{avg}\sqrt{2N})
\end{equation}
were $\sigma_{E}$, $E_{avg}$, and $N$ are the standard deviation, average value, and number of entries for the energy distribution whose resolution is being calculated, respectively.

\subsection{First-order correction using energy deposit in back of calorimeter}

\begin{figure}
    \centering
    \begin{subfigure}{\linewidth}
        \hspace*{-0.5in}
        \includegraphics[scale=0.6]{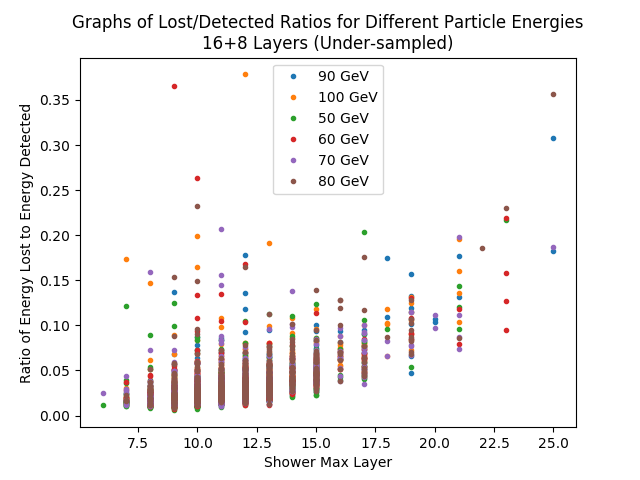}
        \hspace*{-0.8cm}
        \includegraphics[scale=0.6]{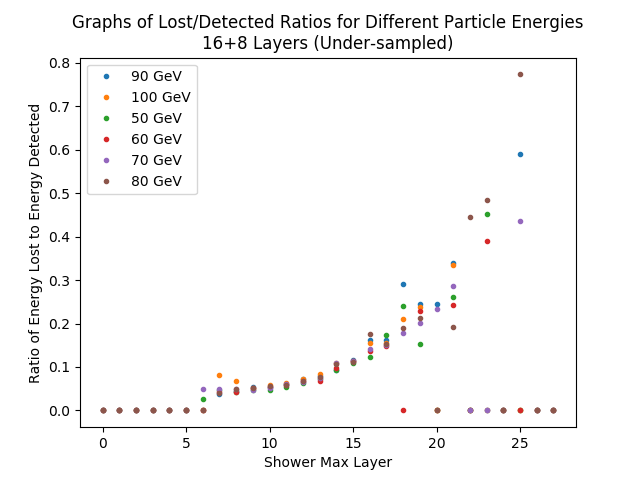}
        \caption{}
        \label{fig:showerMaxTrend}
    \end{subfigure}
    \begin{subfigure}{\linewidth}
        \hspace*{-0.5in}
        \includegraphics[scale=0.6]{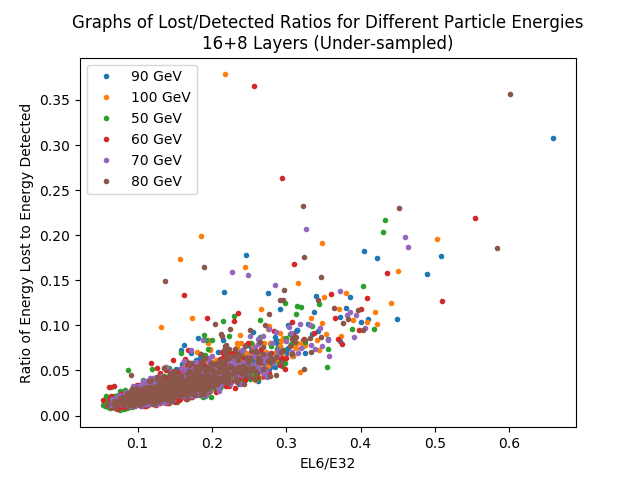}
        \hspace*{-0.8cm}
        \includegraphics[scale=0.6]{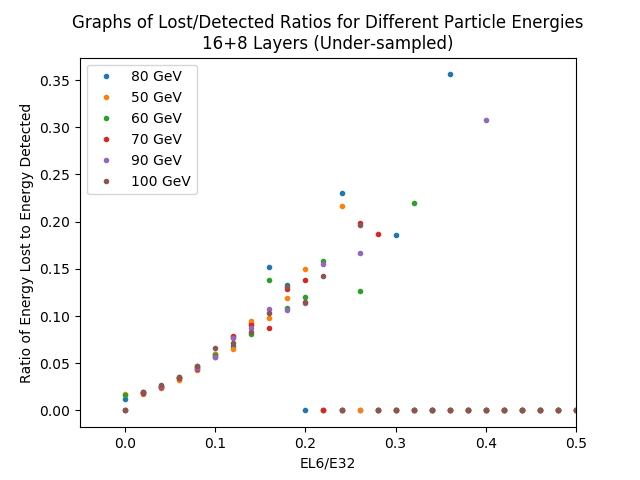}
        \caption{}
        \label{fig:EL6Trend}
    \end{subfigure}
    \caption{First-order correlations between particle shower properties and energy loss. The consistency of these trends across energy levels makes them useful for providing a first-order energy correction. (a) Correlation between the fraction of the energy which was lost and the first layer of shower max for 50, 60, 70, 80, 90, and 100 GeV electrons. Shower max was defined as the three consecutive layers with the highest combined energy deposition. The first graph shows individual events (1000 events per energy level). The second graph shows the average fraction of the energy lost for each energy level. (b) Correlation between the fraction of the energy which was lost and the energy deposited in the last six layers of the calorimeter. Similar trends also exist using the last five layers, seven layers, etc. The first graph shows individual events (1000 events per energy level). The second graph shows the average fraction of the energy lost for each energy level. Average loss was calculated using bins of size 0.02.}
    \label{fig:showerMaxAndEL6Trends}
\end{figure}

\begin{figure}
    \centering
    \begin{subfigure}{\linewidth}
        \hspace*{-0.5in}
        \includegraphics[scale=0.5]{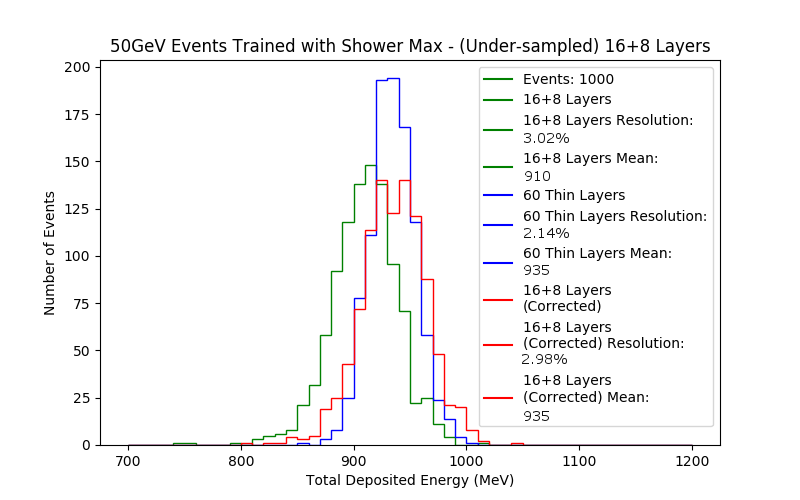}
        \hspace*{-0.8cm}
        \includegraphics[scale=0.5]{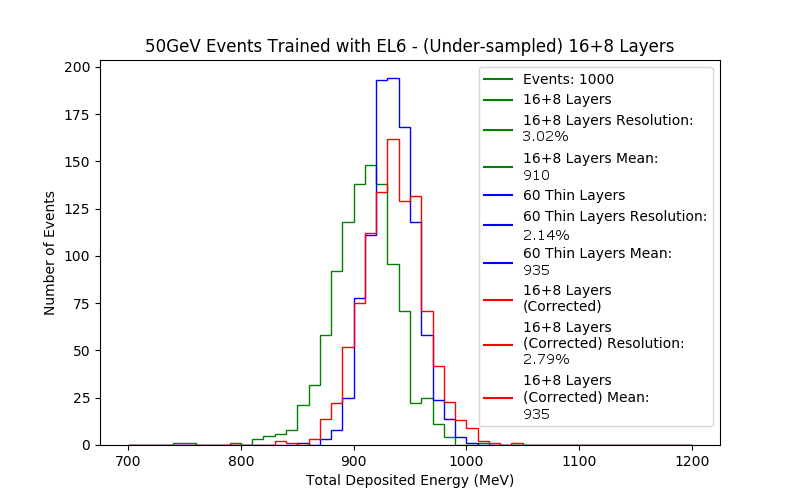}
        \caption{}
        \label{fig:50GeVFirstOrder}
    \end{subfigure}
    \begin{subfigure}{\linewidth}
        \hspace*{-0.5in}
        \includegraphics[scale=0.5]{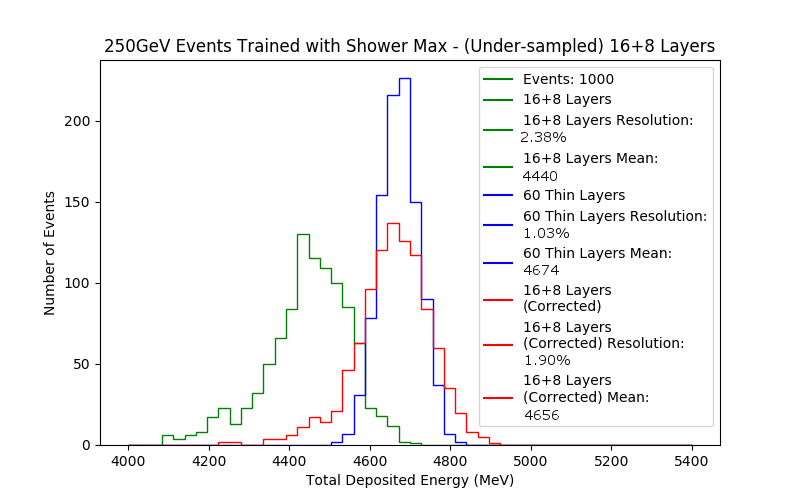}
        \hspace*{-0.8cm}
        \includegraphics[scale=0.5]{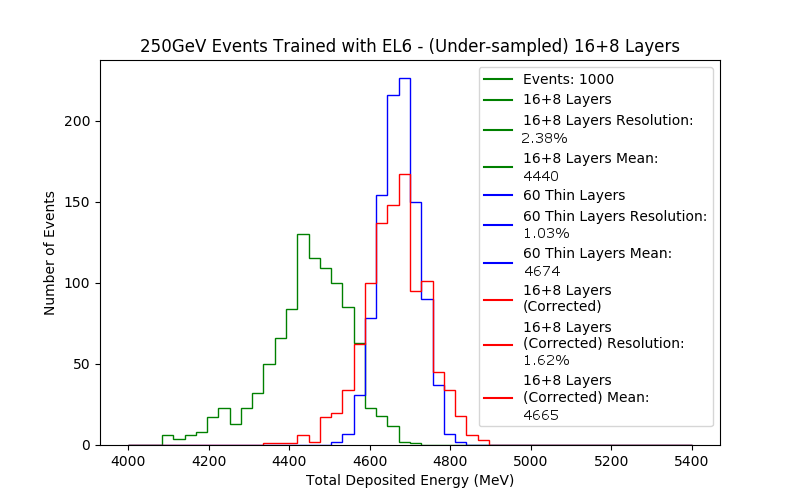}
        \caption{}
        \label{fig:250GeVFirstOrder}
    \end{subfigure}
    \caption{(a) Performance of the shower max and EL6 first order corrections on 50 GeV electron events. Both corrections improve the resolution and average accuracy significantly, but EL6 works slightly better. (b) Performance of the shower max and EL6 first order corrections on 50 GeV electron events. Both corrections improve the average accuracy significantly and improve the resolution somewhat. However, the EL6 correction improves the resolution far more than the shower max correction.}
    \label{fig:discrete_discrete}
\end{figure}

Preliminary energy correction was performed by considering properties of particle showers which correlated with the fraction of total electron energy which was lost. Properties were chosen such that the correlation remained nearly constant as particle energy changed. We evaluated the performance of first-order correction methods using two energy-invariant correlations: (1) the layer at which shower max began and (2) the fraction of the total detected energy measured in the last six layers (EL6) of the calorimeter. Shower max was defined as the three consecutive layers with the highest combined energy deposition. The correlations between these two properties and the fraction of the energy lost are shown in figures \ref{fig:showerMaxTrend} and \ref{fig:EL6Trend}, respectively. To use these correlations to provide correction energy, we computed the correlation for a given electron energy and used it to provide a correction to events of a different energy.

We improved the performance of the first-order correction by computing first-order correction energies for a large number of different electron energies and weighting them with the equation:

\begin{equation}
    \frac{1}{(E_{avg} - E_{32})^{2}}
    \label{FirstOrderWeighting}
\end{equation}
where $E_{avg}$ is the average energy deposited in 32 layers for a given training data set and $E_{32}$ is the energy in 32 layers for the event being corrected. This helped to eliminate the small energy-dependent fluctuations in the first-order correlations which occurred at very low and high electron energies.

Figures \ref{fig:50GeVFirstOrder} and \ref{fig:250GeVFirstOrder} show examples of the weighted first-order correction used on 50 and 250 GeV events, respectively. Both of the first-order correction methods showed substantial improvement over uncorrected events. However, the EL6 correction exhibited significantly better resolution and average accuracy than the shower max correction, especially at high electron energies.

\subsection{Neural network correction - constant energies at constant angles trained on constant energies}

\begin{table}[]
    \centering
    \begin{tabular}{|c|c|}
        \hline
        Network type & Fully connected \\
        \hline
        Hidden layers & 1 (33 nodes) \\
        \hline
        Free parameters & 1156 \\
        \hline
        Activation function & ReLu \\
        \hline
        Optimizer & Adam \\
        \hline
        Loss & Mean squared error \\
        \hline
        Libraries & Keras with TensorFlow backend \cite{chollet_keras_2015,martin_abadi_tensorflow:_2015} \\
        \hline
    \end{tabular}
    \caption{Neural network configuration used to correct constant electron energies entering the ECal at constant angles when training on constant electron energies}
    \label{tab:discreteEnergy_discreteAngle_discreteTraining}
\end{table}

\begin{figure}
    \centering
    \begin{subfigure}{\linewidth}
        \hspace*{-0.8in}
        \includegraphics[scale=0.5]{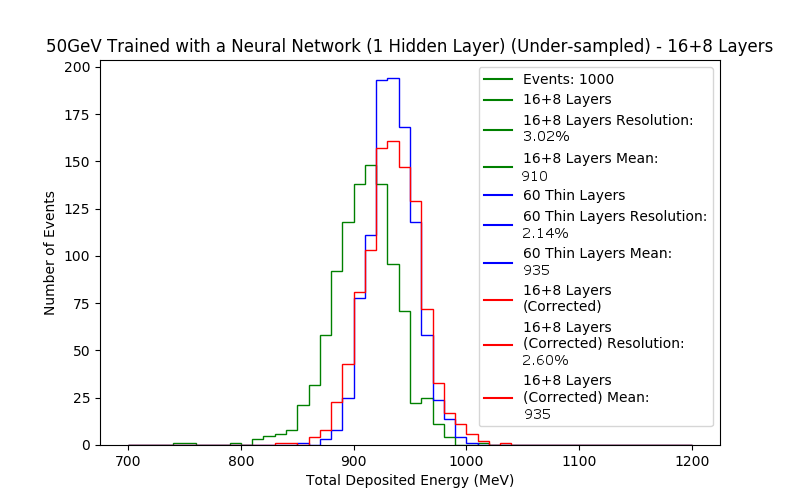}
        \hspace*{-0.5cm}
        \includegraphics[scale=0.5]{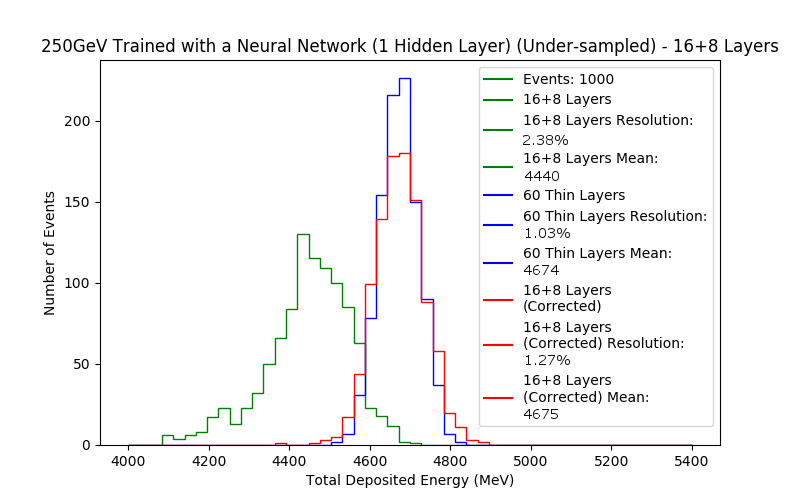}
    \end{subfigure}
    \caption{Performance of the NN correction trained constant energy electron data sets when correcting 50 and 250 GeV electron events. The NN correction works excellently in both cases and significantly outperforms the first-order corrections, especially at high electron energies.}
    \label{fig:50and250GeV_discreteEnergy_discreteAngle_discreteTraining}
\end{figure}

\begin{figure}
    \centering
    \includegraphics[scale=0.8]{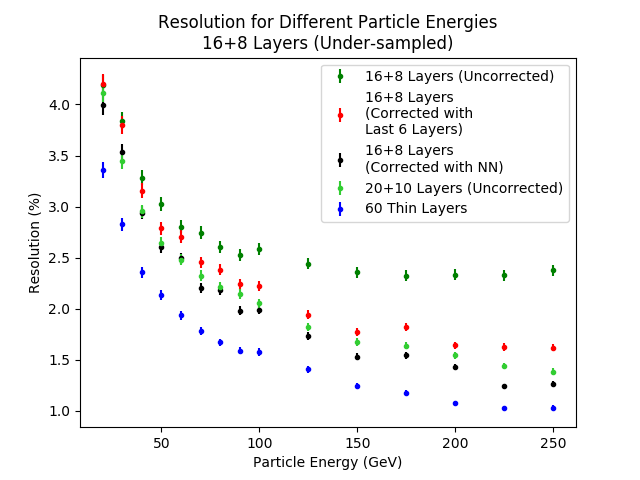}
    \caption{Resolutions at different electron energies for different ECal setups. The 16+8 layer setup corrected using either EL6 or a NN trained on constant electron energies performed comparably well to the uncorrected 20+10 layer setup, and the NN correction consistently performed slightly better than the 20+10 setup, though not always with statistical significance.}
    \label{fig:Resolution_discreteEnergy_discreteAngle_discreteTraining}
\end{figure}

We used a NN to provide a second-order energy correction to electron events with constant energies. Table \ref{tab:discreteEnergy_discreteAngle_discreteTraining} shows the specifications of the NN used. The NN was given the total energy deposited in each of the 32 layers and the total energy in all 32 layers and asked to predict the amount of energy lost in MeV. This allowed for a more thorough analysis of energy loss trends than using a first-order correction.

Correction energy was calculated for each electron energy separately, and the correction energy for a given event was the weighted average of the correction energies for all electron energies other than the one being tested. Unlike the first-order correction, the NN correction was weighted by:

\begin{equation}
    \frac{1}{(E_{avg} - E_{32})^{10}}
    \label{eqn:NNWeighting}
\end{equation}
where $E_{avg}$ is the average energy deposited in 32 layers for a given training data set and $E_{32}$ is the energy in 32 layers for the event being corrected. Weighting by the difference to the tenth power ensured that only the data sets closest in electron energy to the event being corrected would significantly impact the correction. This was necessary because the NN's performance was more energy-dependent than the first-order fit.

The performance of the NN correction on 50 and 250 GeV events is shown in figure \ref{fig:50and250GeV_discreteEnergy_discreteAngle_discreteTraining}. The NN correction provided excellent improvement over uncorrected events and significantly outperformed the first-order correction, especially at high electron energies.

Figure \ref{fig:Resolution_discreteEnergy_discreteAngle_discreteTraining} shows the resolutions which were achieved for different electron energies by using the first-order correction, the NN correction, and 20+10 layers (uncorrected). Both the first-order EL6 and NN corrections performed comparably to the 20+10 design, and NN correction consistently performed slightly better than 20+10.

\subsection{Neural network correction - constant energies at constant angles trained on energy distributions}

\begin{table}[]
    \centering
    \begin{tabular}{|c|c|}
        \hline
        Network type & Fully connected \\
        \hline
        Hidden layers & 8 (33, 24, 16, 12, 8, 6, 4, 2 nodes) \\
        \hline
        Free parameters & 2741 \\
        \hline
        Activation function & ReLu \\
        \hline
        Optimizer & Adam \\
        \hline
        Loss & Mean squared error \\
        \hline
        Libraries & Keras with TensorFlow backend \\
        \hline
    \end{tabular}
    \caption{Neural network configuration used to correct constant electron energies and electron energy distributions entering the ECal at constant angles when training on electron energy distributions}
    \label{tab:discreteEnergy_discreteAngle_distTraining}
\end{table}

\begin{figure}
    \centering
    \begin{subfigure}{\linewidth}
        \hspace*{-0.5in}
        \includegraphics[scale=0.52]{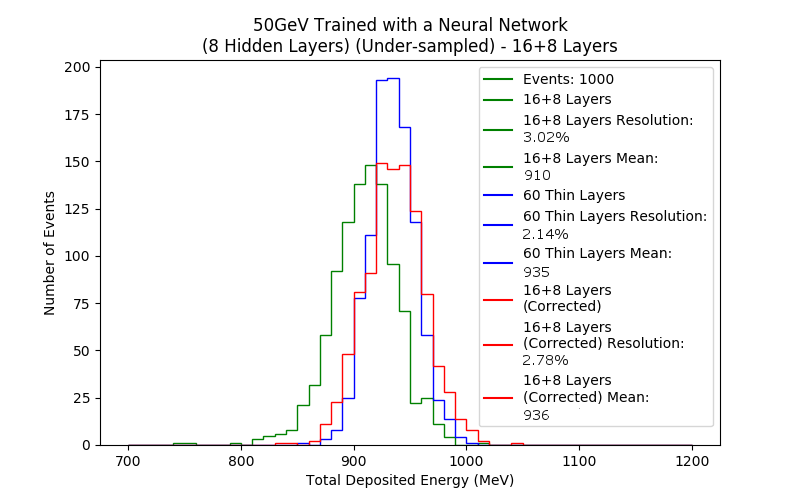}
        \hspace*{-1.0cm}
        \includegraphics[scale=0.52]{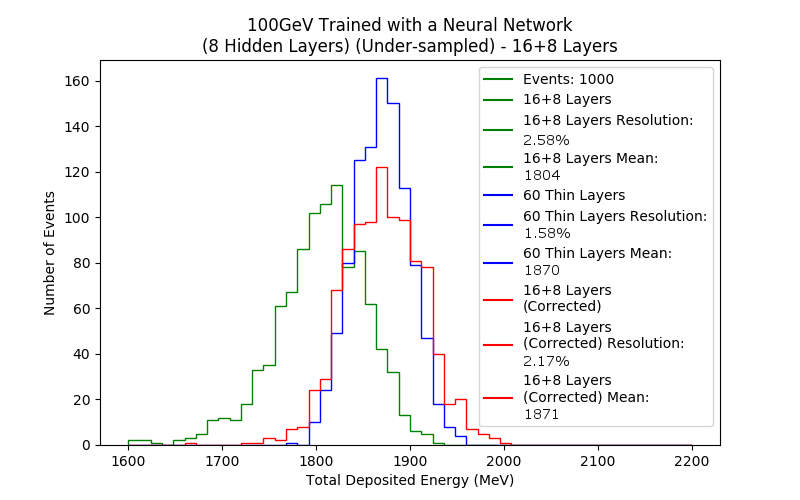}
    \end{subfigure}
    \begin{subfigure}{\linewidth}
        \centering
        \includegraphics[scale=0.6]{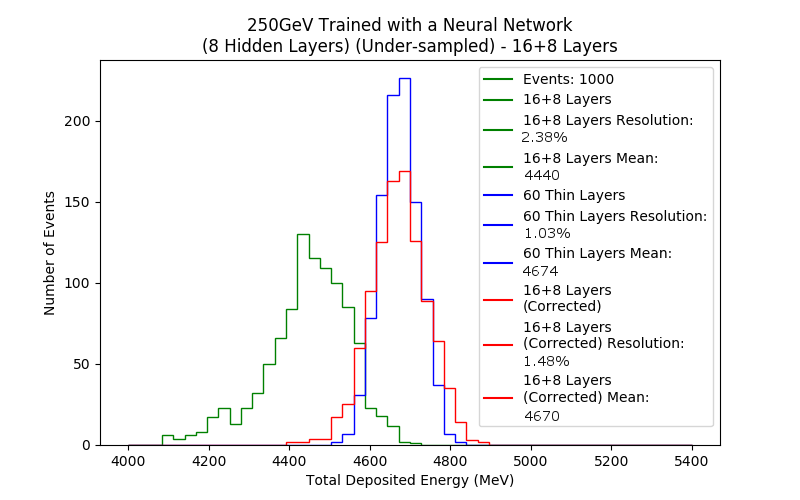}
    \end{subfigure}
    \caption{Performance of the NN correction trained electron energy distribution data sets when correcting 50, 100, and 250 GeV electron events. The energy distribution NN correction works well in all cases and significantly outperforms the first-order corrections, especially at high electron energies. However, it does not outperform the NN trained on constant electron energies.}
    \label{fig:50and100GeV_discreteEnergy_discreteAngle_distTraining}
\end{figure}

\begin{figure}
    \centering
    \includegraphics[scale=0.8]{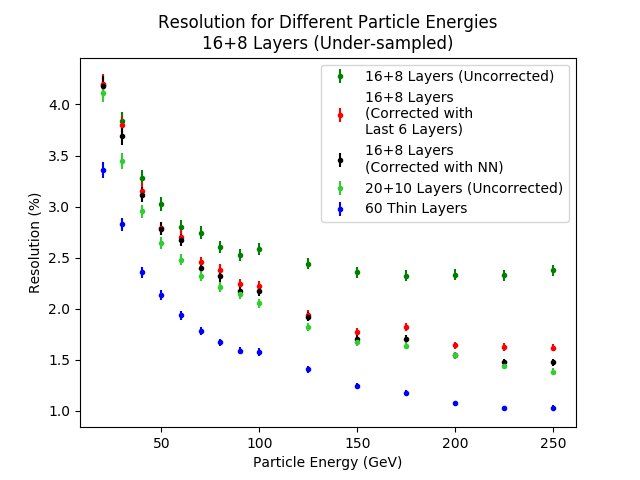}
    \caption{Resolutions at different electron energies for different ECal setups. The 16+8 layer setup corrected using either EL6 or a NN performed comparably well to the uncorrected 20+10 layer setup. In this case, the NN was trained on electron energy distributions (one 10-300GeV and one 20-300GeV) instead of constant electron energies and never outperformed the 20+10 design.}
    \label{fig:Resolutions_discreteEnergy_discreteAngle_distTraining}
\end{figure}

To make our model more realistic, we trained a NN on electron events with a distribution of energies. Some distributions were between 10 and 300 GeV, while others were between 20 and 300 GeV. We then evaluated the NN correction's performance to compare with the constant-energy NN's performance. The neural network setup is shown in table \ref{tab:discreteEnergy_discreteAngle_distTraining}.

Figure \ref{fig:50and100GeV_discreteEnergy_discreteAngle_distTraining} shows total energy deposits and resolutions for 16+8 uncorrected, 16+8 corrected with the distribution-trained NN, and 60 thin layers (ideal). The NN consistently corrected the mean of the distribution and significantly improved the resolution, especially at high energies. The 16+8 corrected resolution was comparable to that of the 20+10 design but was slightly low in most cases (see figure \ref{fig:Resolutions_discreteEnergy_discreteAngle_distTraining}). This is in contrast to when the NN was trained on constant electron energies, in which case the 16+8 corrected resolution was lower than the 20+10 resolution (see figure \ref{fig:Resolution_discreteEnergy_discreteAngle_discreteTraining}).

\begin{figure}
    \centering
    \begin{subfigure}{\linewidth}
        \hspace*{-0.5in}
        \includegraphics[scale=0.52]{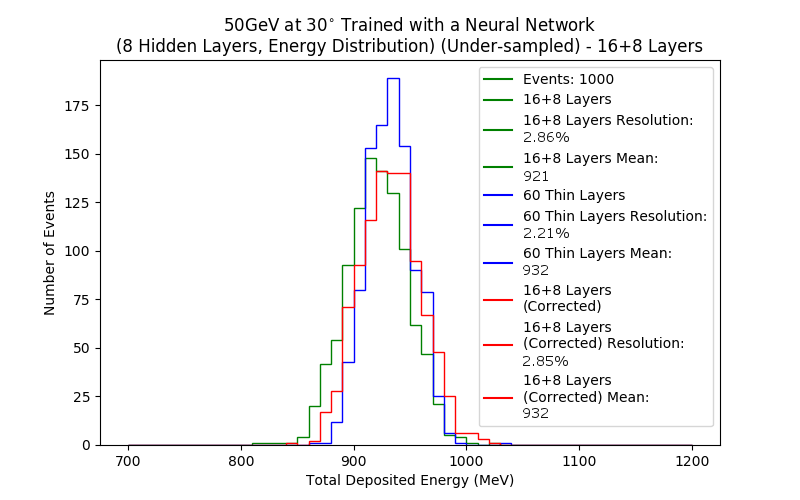}
        \hspace*{-1.0cm}
        \includegraphics[scale=0.52]{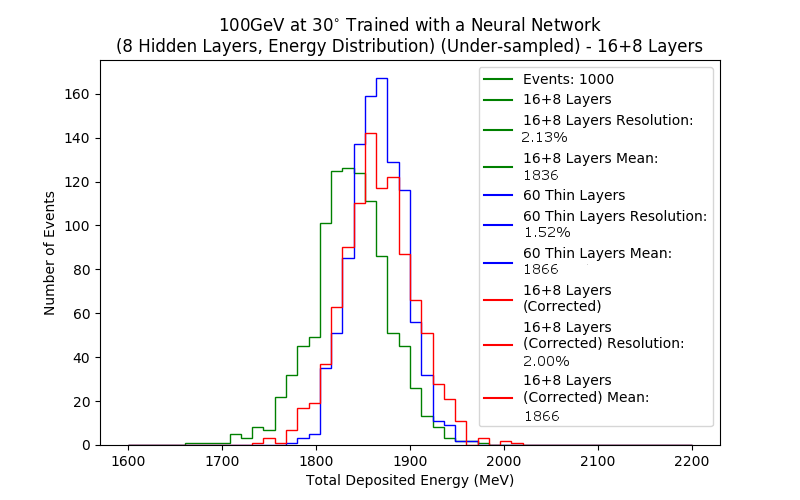}
    \end{subfigure}
    \caption{Performance of the NN correction trained electron energy distribution data sets when correcting 50 and 100 GeV electron events incident in the calorimeter at 30$^{\circ}$ from the normal. The energy distribution NN correction works well in both cases. However, the improvement over the uncorrected events is smaller than for events incident normal to the ECal, as there is less leakage at steeper angles.}
    \label{fig:50and100GeV_30degrees_discreteEnergy_discreteAngle_distTraining}
\end{figure}

In order to preliminarily evaluate how the performance of the neural network and the behavior of the particle showers changed as the angle of incidence changed, we trained a NN on distributions of energy between 20 and 300 GeV incident on the ECal at 30$^{\circ}$ from the normal. We tested the NN correction on constant energy electron data sets for comparison with events which were normal to the detector. At 30$^{\circ}$, the NN still corrected the average deposited energy and improved the resolution (see figure \ref{fig:50and100GeV_30degrees_discreteEnergy_discreteAngle_distTraining}). However, the NN did not provide as large of an improvement over the uncorrected energies as at 0$^{\circ}$ from normal. This is because the electron passes through more detector material resulting in less energy leakage. Less leakage means that there is less for the NN to correct and the improvement from the correction is smaller.

\subsection{Neural network correction - energy distributions at constant angles}

\begin{figure}
    \centering
    \begin{subfigure}{\linewidth}
        \hspace*{-0.5in}
        \includegraphics[scale=0.6]{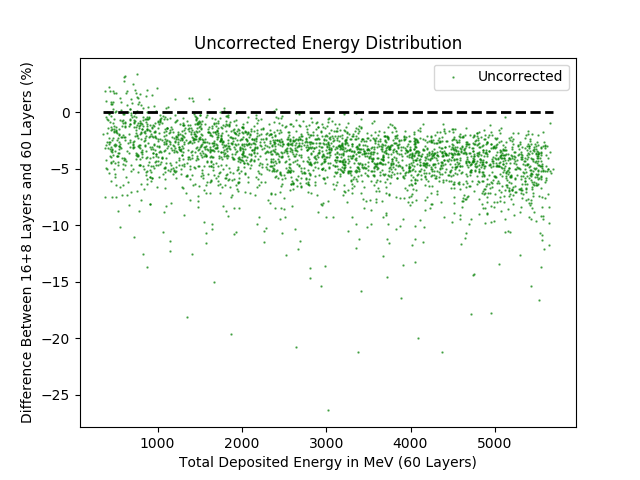}
        \hspace*{-0.5cm}
        \includegraphics[scale=0.6]{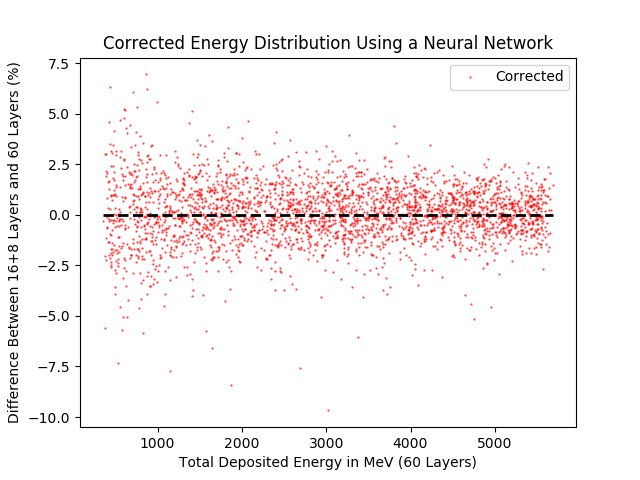}
        \caption{}
        \label{fig:Scatterplots_distEnergy_discreteAngle}
    \end{subfigure}
    \begin{subfigure}{\linewidth}
        \hspace*{-0.5in}
        \includegraphics[scale=0.52]{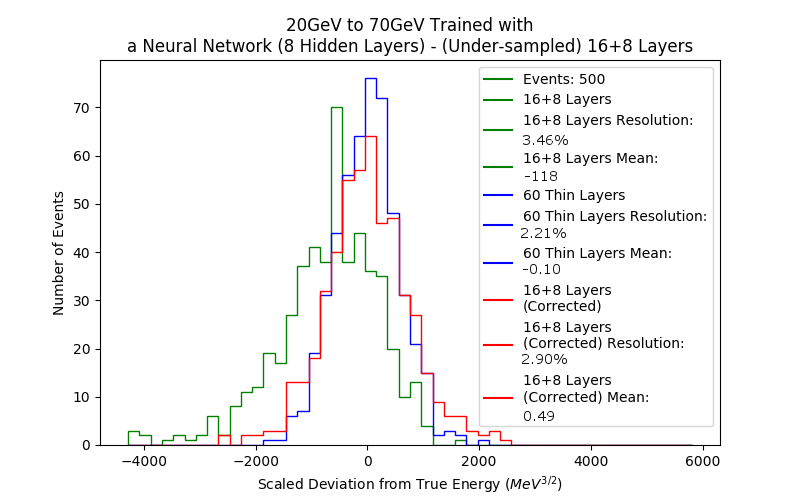}
        \hspace*{-1.0cm}
        \includegraphics[scale=0.52]{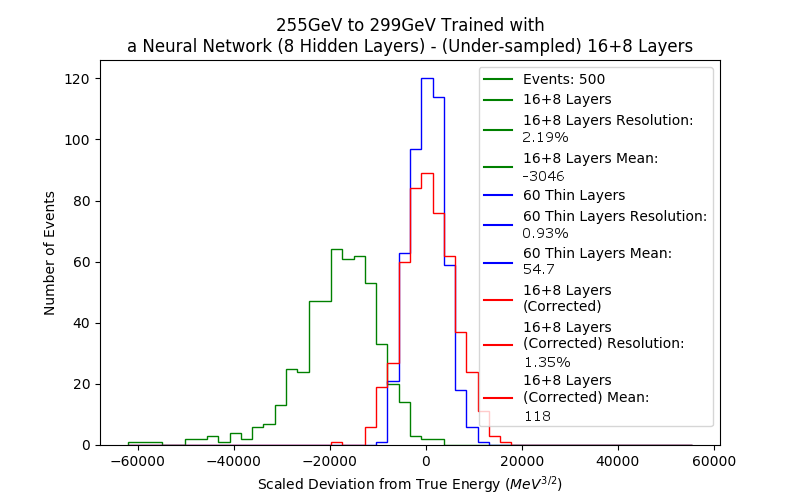}
        \caption{}
        \label{fig:20and255GeV_distEnergy_discreteAngle}
    \end{subfigure}
    \caption{(a) Percent deviation from the ideal energy deposit for 3000 events with energies between 20 and 300 GeV before and after the energy-distribution-trained NN correction is applied. There is a higher percentage leakage at higher electron energies, but the algorithm consistently corrects the average of the distribution accurately. (b) Total energy deposits and resolutions for 16+8 uncorrected, 16+8 corrected, and 60 thin layers (ideal). The correction accurately predicts the mean of the distribution and makes the resolution much lower for higher energies.}
\end{figure}

\begin{figure}
    \centering
    \includegraphics[scale=0.6]{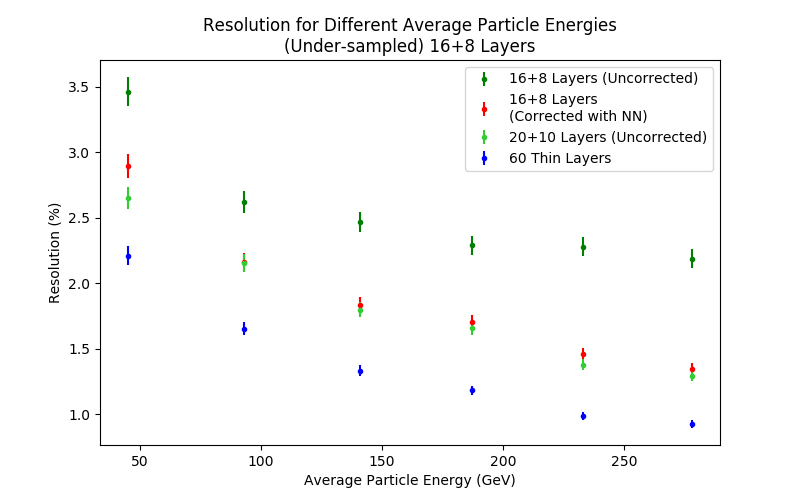}
    \caption{Resolutions for a distribution of electron energies for different ECal setups. Each point represents the average resolution across 500 events. The 16+8 layer setup was corrected using a NN trained on two energy distributions, one 10-300GeV and one 20-300GeV. The corrected 16+8 design performed statistically equivalently to the 20+10 design in all cases except the lowest energy events.}
    \label{fig:Resolutions_distEnergy_discreteAngle}
\end{figure}

To further make the model reflect the ECal's experimental conditions, we evaluated the ability of the NN trained on a uniformly random distribution of energies to evaluate another uniformly random distribution of energies. Table \ref{tab:discreteEnergy_discreteAngle_distTraining} shows the NN's specifications.

To be able to compare energy deposits and resolution values for different electron energies, we computed the average deposition which would be expected for each electron energy if we were to have a large number of electrons of that energy. This energy is approximately given by:

\begin{equation}
    E_{avg} \approx 0.01869E_{e^{-}}
\end{equation}
for electrons fired normal to the calorimeter's surface, where $E_{e^{-}}$ is the total energy of the electron. These values were used to normalize each distribution for plotting in a histogram. Each energy value was also raised to the power of 1.5 before normalizing, because events within each bin needed to have comparable resolutions, and the resolution decreases with the square root of the energy.

Figure \ref{fig:Scatterplots_distEnergy_discreteAngle} shows the percent deviation from the ideal energy deposit for 3000 events with energies between 20 and 300 GeV before and after the NN correction is applied. Figure \ref{fig:20and255GeV_distEnergy_discreteAngle} shows the resolutions and total energy deposits for ranges of low- and high-energy particles. The NN showed an excellent ability to correct the mean value of the energy deposit for all particle energies. It was able to drastically reduce the measurement resolution, especially at high electron energies.

Figure \ref{fig:Resolutions_distEnergy_discreteAngle} shows the resolutions achieved by the NN correction for different electron energies and different ECal designs. The 16+8 with NN correction design consistently provided a large improvement in resolution over 16+8 uncorrected design, with resolutions highly comparable to the 20+10 design resolution for energies above 50 GeV.

\subsection{Neural network correction - constant energies at variable angles}

\begin{table}[]
    \centering
    \begin{tabular}{|c|c|}
        \hline
        Network type & Fully connected \\
        \hline
        Hidden layers & 8 (34, 24, 16, 12, 8, 6, 4, 2 nodes) \\
        \hline
        Free parameters & 2833 \\
        \hline
        Activation function & ReLu \\
        \hline
        Optimizer & Adam \\
        \hline
        Loss & Mean squared error \\
        \hline
        Libraries & Keras with TensorFlow backend \\
        \hline
    \end{tabular}
    \caption{Neural network configuration used to correct constant electron energies entering the ECal at variable angles when training on constant electron energies entering the ECal at variable angles}
    \label{tab:discreteEnergy_distAngle_distTraining}
\end{table}

%\begin{figure}
%    \centering
%    \includegraphics[scale=0.7]{DiscreteEnergyDistAngleNNGraphs/50GeV/testNN_Distribution_Uncorrected.png}
%    \includegraphics[scale=0.7]{DiscreteEnergyDistAngleNNGraphs/50GeV/testNN_Distribution_Corrected.png}
%    \caption{Caption}
%    \label{fig:Scatterplots_50GeV_discreteEnergy_distAngle}
%\end{figure}

\begin{figure}
    \centering
    \begin{subfigure}{\linewidth}
        \hspace*{-0.5in}
        \includegraphics[scale=0.6]{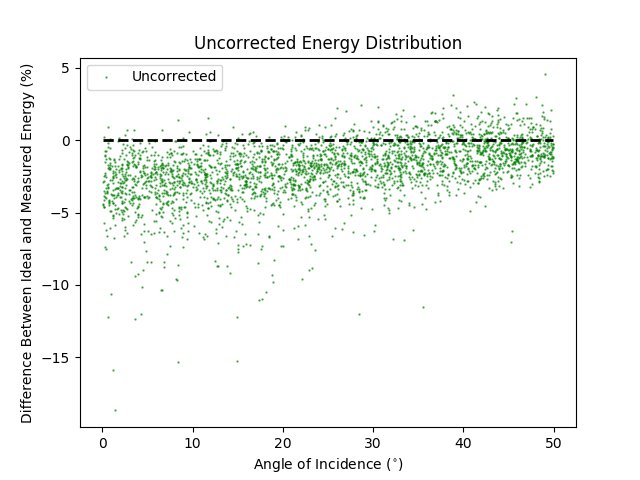}
        \hspace*{-0.5cm}
        \includegraphics[scale=0.6]{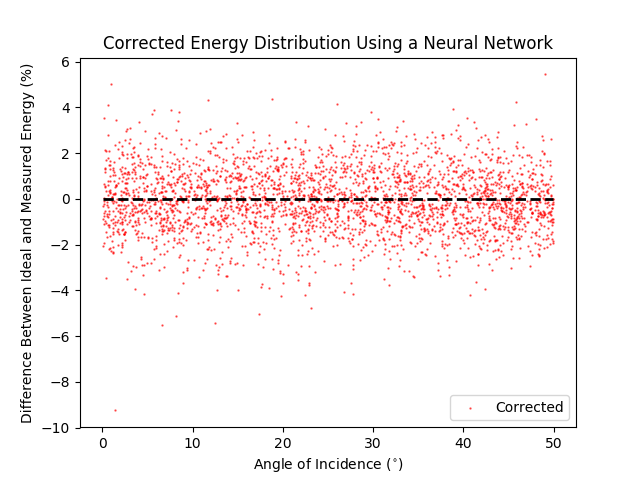}
        \caption{}
        \label{fig:Scatterplots_100GeV_discreteEnergy_distAngle}
    \end{subfigure}
    \begin{subfigure}{\linewidth}
        \hspace*{-0.75in}
        \includegraphics[scale=0.52]{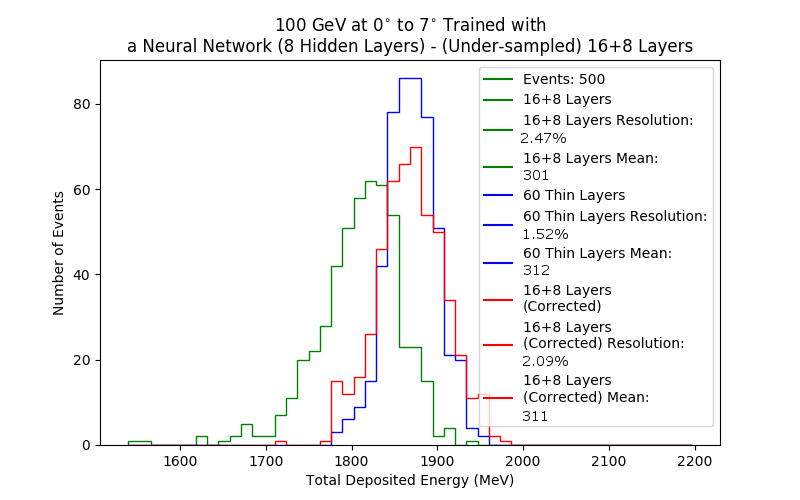}
        \hspace*{-1.0cm}
        \includegraphics[scale=0.52]{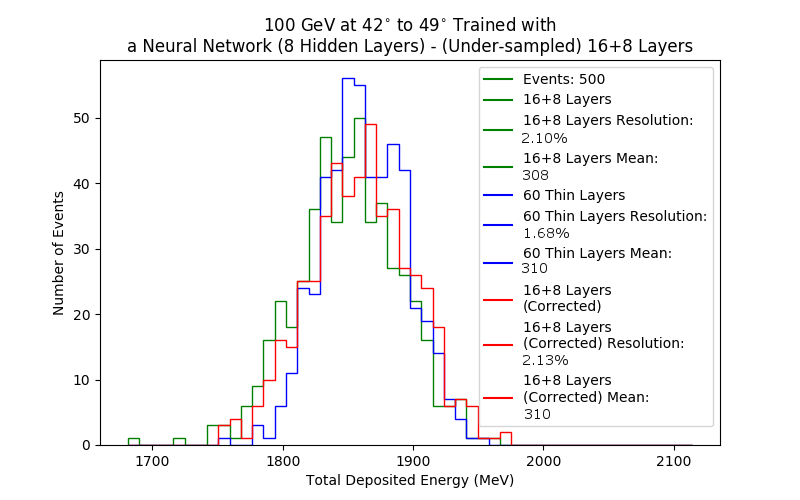}
        \caption{}
        \label{fig:0and42_100GeV_discreteEnergy_distAngle}
    \end{subfigure}
    \caption{(a) Percent deviation from the ideal energy deposit for 3000 events at 100 GeV incident at random angles between 0$^{\circ}$ and 50$^{\circ}$ before and after the NN correction is applied. The average accuracy and resolution improve as the angle of incidence increases. This is because the electron passes through more material, so there is less leakage. The NN trained on 100GeV events incident at random angles between 0$^{\circ}$ and 50$^{\circ}$ consistently corrects the average value for all angles. (b) Total energy deposits and resolutions for 16+8 uncorrected, 16+8 corrected, and 60 thin layers (ideal) at a range of shallow and steep angles. The NN correction corrects the average energy deposit in both cases but only improves the resolution at shallow angles, as the resolution is already low at steep angles.}
\end{figure}

%\begin{figure}
%    \centering
%    \includegraphics[scale=0.7]{DiscreteEnergyDistAngleNNGraphs/50GeV/Hist_3000_0_a_Neural_Network_(8_Hidden_Layers)_32.png}
%    \includegraphics[scale=0.7]{DiscreteEnergyDistAngleNNGraphs/50GeV/Hist_3000_40_a_Neural_Network_(8_Hidden_Layers)_32.png}
%    \caption{Caption}
%    \label{fig:0and40_50GeV_discreteEnergy_distAngle}
%\end{figure}

\begin{figure}
    \centering
    \includegraphics[scale=0.6]{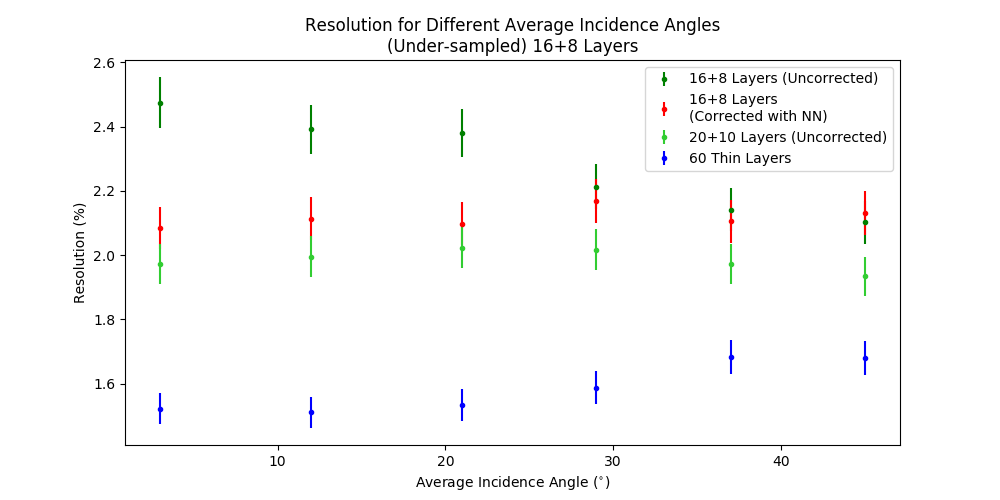}
    \caption{Resolutions for 100GeV events at a distribution of incidence angles for different ECal setups. Each point represents the average resolution across 500 events. The 16+8 layer setup was corrected using a NN trained on 100GeV events incident at random angles between 0$^{\circ}$ and 50$^{\circ}$ (same range as for testing). The corrected 16+8 design performed statistically equivalently to the 20+10 design for shallow angles and performed worse at steep angles. As the angle increases, the resolution of the 16+8-layer design gets better and the resolution of the 60 thin layers design gets worse. This is because the leakage decreases and the sampling gets coarser at high angles. In the 16+8 design, leakage is the dominant factor in determining error, whereas sampling coarseness is the dominant factor for the 60-layer design.}
    \label{fig:Resolutions_discreteEnergy_distAngle}
\end{figure}

We trained a NN on single electron events at constant energies entering the calorimeter at variable angles between 0$^{\circ}$ and 50$^{\circ}$ to more clearly evaluate the NN correction's performance at different angles of incidence. Here, we present results for 100 GeV electrons, but similar results were also obtained with 50 GeV electrons. The specifications for the NN used are shown in table \ref{tab:discreteEnergy_distAngle_distTraining}.

Figure \ref{fig:Scatterplots_100GeV_discreteEnergy_distAngle} shows the percent deviation from the ideal energy deposit for 3000 events at 100 GeV before and after the NN correction is applied as a function of incidence angle. Figure \ref{fig:0and42_100GeV_discreteEnergy_distAngle} shows the resolutions and total energy deposits for a range of steep and shallow angle events. The correction consistently shifts the mean average energy such that it agrees with the average true energy to within 1\%. Very high-loss events are also corrected to be far closer to the center of the distribution. This results in a substantial resolution decrease at low angles. However, the average energy loss and resolution for the uncorrected events both decrease as the angle of incidence increases, so the NN correction becomes unnecessary at high angles.

Figure \ref{fig:Resolutions_discreteEnergy_distAngle} shows the resolutions achieved at different angles of incidence for different detector setups at 100 GeV. The NN correction significantly improved the resolution of the 16+8 layer design and made it comparable to the resolution for the 20+10 layer design for angles below about 25$^{\circ}$. This is in agreement with figure \ref{fig:Scatterplots_100GeV_discreteEnergy_distAngle}.

%%% IN PROGRESS %%%

\subsection{Neural network correction - energy distributions at variable angles}

\begin{table}[]
    \centering
    \begin{tabular}{|c|c|}
        \hline
        Network type & Fully connected \\
        \hline
        Hidden layers & 1 (34 nodes) \\
        \hline
        Free parameters & 1225 \\
        \hline
        Activation function & ReLu \\
        \hline
        Optimizer & Adam \\
        \hline
        Loss & Mean squared error \\
        \hline
        Libraries & Keras with TensorFlow backend \\
        \hline
    \end{tabular}
    \caption{Neural network configuration used to correct variable electron energies entering the ECal at variable angles when training on variable electron energies entering the ECal at variable angles}
    \label{tab:distEnergy_distAngle}
\end{table}

\begin{figure}
    \centering
    \begin{subfigure}{\linewidth}
        \hspace*{-0.5in}
        \includegraphics[scale=0.43]{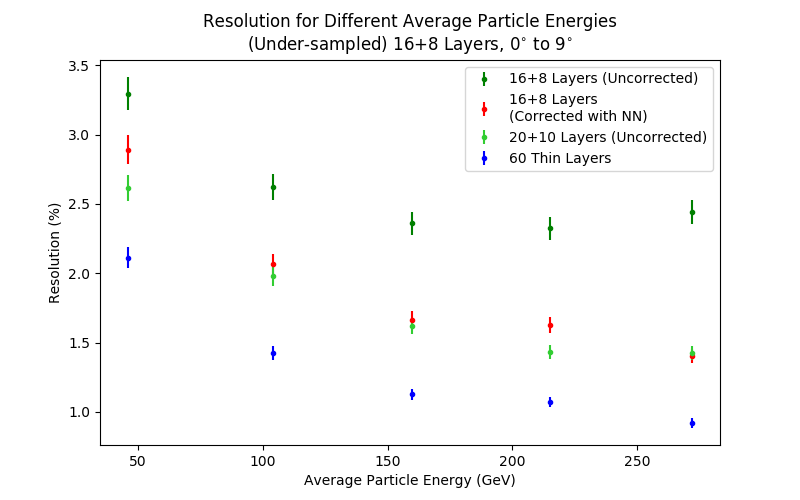}
        \hspace*{-0.8cm}
        \includegraphics[scale=0.43]{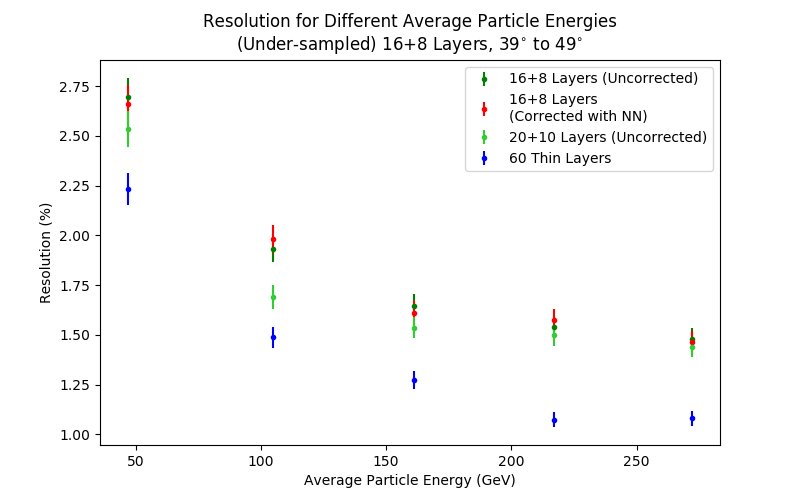}
        \caption{}
        \label{fig:Resolutions_distEnergy_distAngle}
    \end{subfigure}
    \begin{subfigure}{\linewidth}
        \hspace*{-0.1in}
        \includegraphics[scale=0.45]{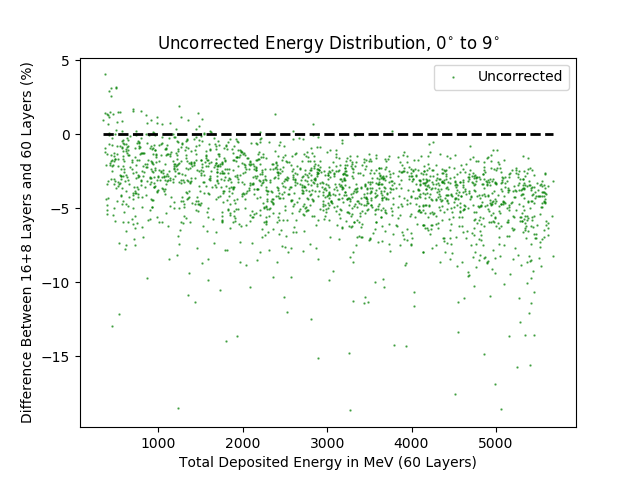}
        \hspace*{-0.5cm}
        \includegraphics[scale=0.45]{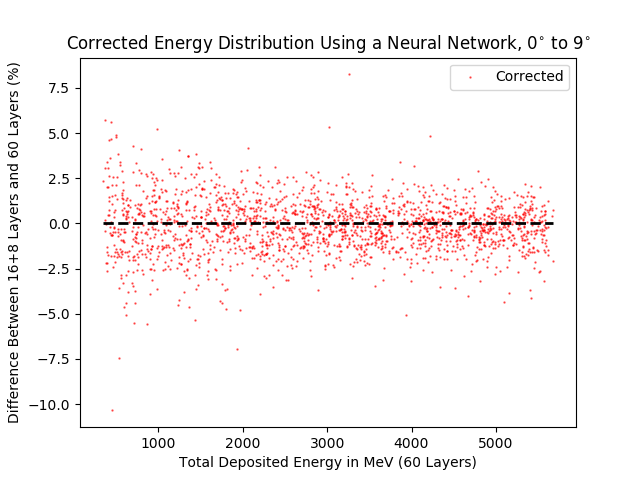}
        \caption{}
        \label{fig:Scatterplots_LowAngle_distEnergy_distAngle}
    \end{subfigure}
    \begin{subfigure}{\linewidth}
        \hspace*{-0.1in}
        \includegraphics[scale=0.45]{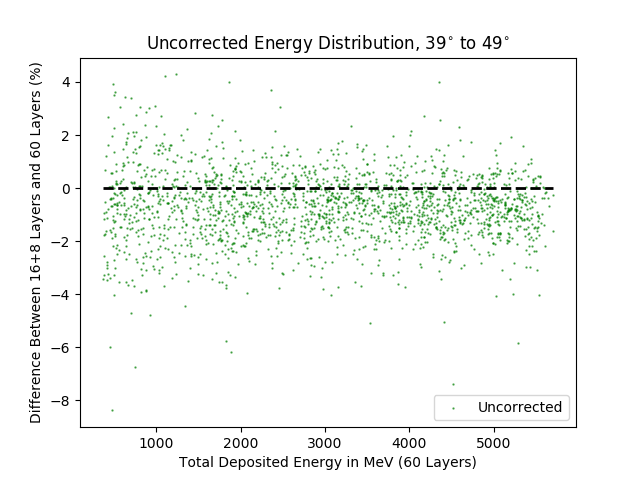}
        \hspace*{-0.5cm}
        \includegraphics[scale=0.45]{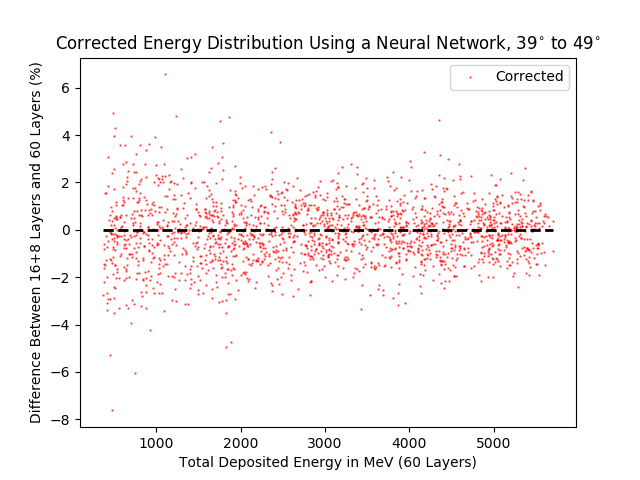}
        \caption{}
        \label{fig:Scatterplots_HighAngle_distEnergy_distAngle}
    \end{subfigure}
    \caption{\setstretch{0.5}(a) Resolutions for a distribution of electron energies incident at a range of shallow and steep angles for different ECal setups. Each point represents the average resolution across 400 events. The 16+8 layer setup was corrected using a NN trained on electrons with energies between 20 and 300GeV and incident at random angles between 0$^{\circ}$ and 50$^{\circ}$ (same ranges as for testing). The corrected 16+8 design performed comparably to the 20+10 design for shallow angles across all energies. The corrected 16+8 design did not perform better than the uncorrected 16+8 design at steep angles. This is because the leakage drops significantly as angle increases. (b) Percent deviation from true electron energy for a range of shallow angles (0$^{\circ}$-9$^{\circ}$) before and after the NN correction. The NN correction consistently corrects the mean energy and makes the distribution narrower for all electron energies. (c) Percent deviation from true electron energy for a range of steep angles (39$^{\circ}$-49$^{\circ}$) before and after the NN correction. The NN correction consistently corrects the mean energy for all energies.}
    \label{fig:distEnergy_distAngle}
\end{figure}

We made the model even more realistic by testing the performance of the NN correction on data sets in which the electron energies (between 20 and 300 GeV) and incidence angles (between 0$^{\circ}$ and 50$^{\circ}$) were both determined uniformly randomly. The NN specifications are shown in table \ref{tab:distEnergy_distAngle}. It was determined that using 8 hidden layers did not significantly improve the results and made the algorithm take far longer to run, so 1 hidden layer was used instead.

Figure \ref{fig:Resolutions_distEnergy_distAngle} shows the resolutions achieved when using very low- and high-angle events. Figures \ref{fig:Scatterplots_LowAngle_distEnergy_distAngle} and \ref{fig:Scatterplots_HighAngle_distEnergy_distAngle} show the percent deviation from the ideal energy deposit for 2000 events, at low and high incidence angles respectively, before and after the NN correction was applied. The NN corrected the average energy and resolution very well at shallow angles where the energy loss is the highest. The NN still corrected the mean at steep angles but does not improve the resolution, since the resolution is already very low before the correction at steep angles. Similar trends were observed for intermediate angles.

\subsection{Neural network correction - preliminary implementation in modified full SiD model}

\begin{table}[]
    \centering
    \begin{tabular}{|c|c|}
        \hline
        Network type & Fully connected \\
        \hline
        Hidden layers & 1 (59 nodes) \\
        \hline
        Free parameters & 3600 \\
        \hline
        Activation function & ReLu \\
        \hline
        Optimizer & Adam \\
        \hline
        Loss & Mean squared error \\
        \hline
        Libraries & Keras with TensorFlow backend \\
        \hline
    \end{tabular}
    \caption{Neural network configuration used to correct distributions of electron energies in a modified SiD ECal design fired at random energies (20-300GeV) and angles ($\theta$ = 0$^{\circ}$-45$^{\circ}$, $\phi$ = 0$^{\circ}$-360$^{\circ}$) using a modified version of the full SiD simulation without separating the modules of events with deposits in multiple modules.}
    \label{tab:discreteEnergy_FullSiD_GroupedModules}
\end{table}

\begin{figure}
    \centering
    \begin{subfigure}{\linewidth}
        \hspace*{-0.5in}
        \includegraphics[scale=0.4]{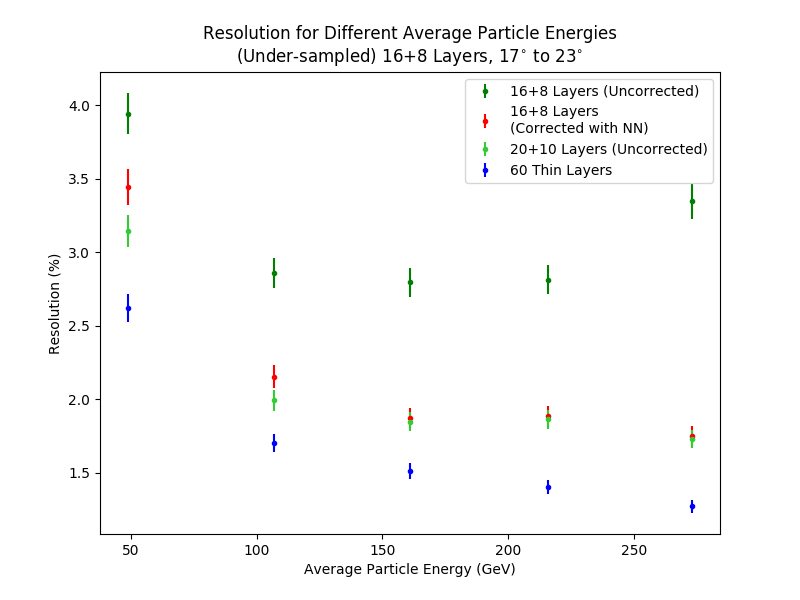}
        \hspace*{-0.8cm}
        \includegraphics[scale=0.4]{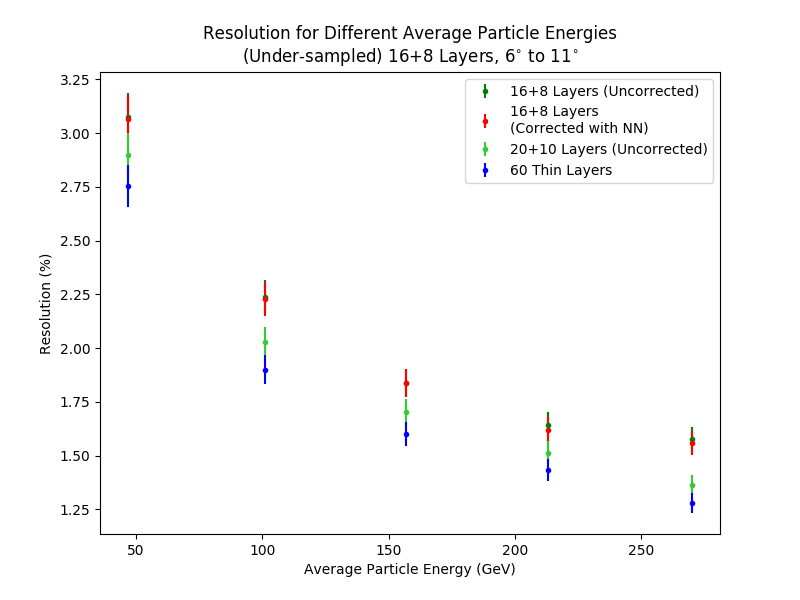}
        \caption{}
        \label{fig:FullSiD_Resolutions}
    \end{subfigure}
    \begin{subfigure}{5cm}
        \hspace*{-1.4in}
        \includegraphics[scale=0.45]{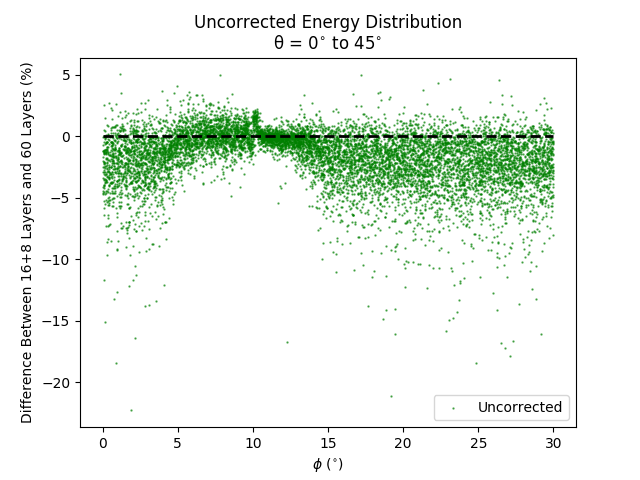}
        \caption{}
        \label{fig:FullSiD_Scatterplots_Phi}
    \end{subfigure}
    \begin{subfigure}{5cm}
        \hspace*{-0.5cm}
        \includegraphics[scale=0.45]{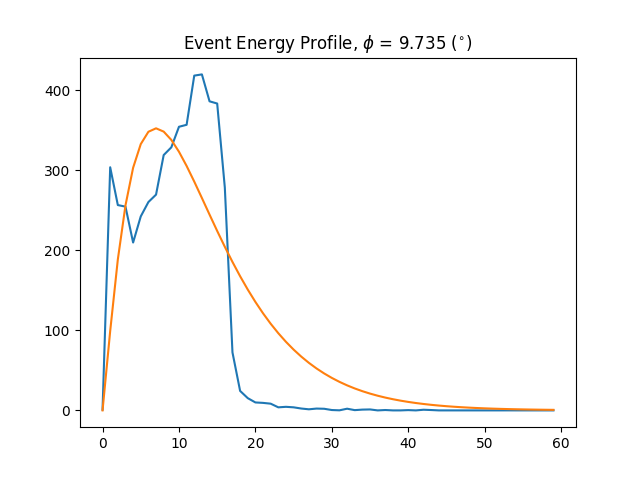}
        \caption{}
        \label{fig:FullSiD_DistortedProfile}
    \end{subfigure}
    \begin{subfigure}{\linewidth}
        \hspace*{-0.1in}
        \includegraphics[scale=0.45]{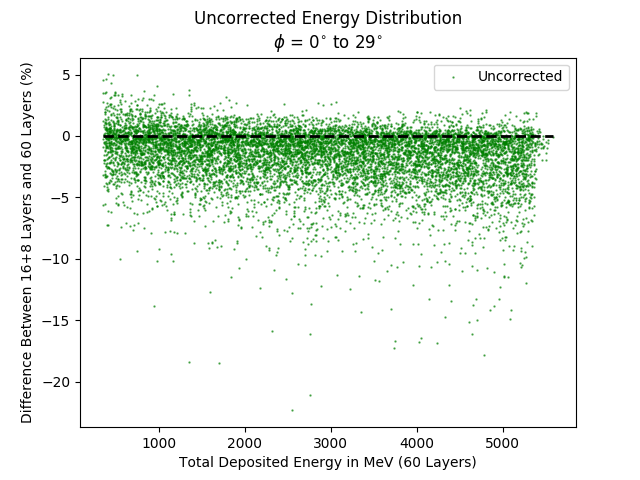}
        \hspace*{-0.5cm}
        \includegraphics[scale=0.45]{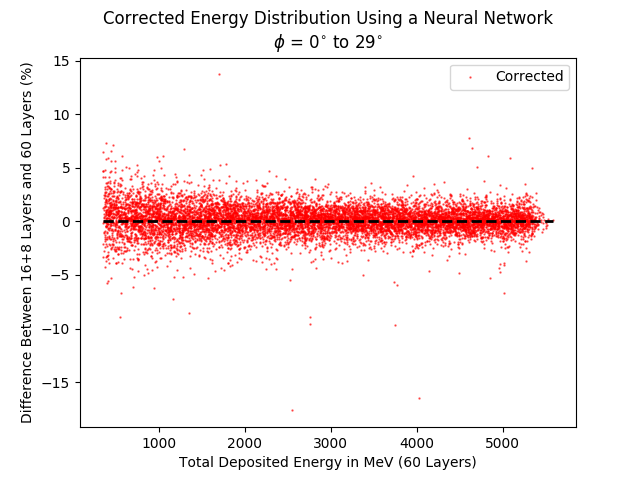}
        \caption{}
        \label{fig:FullSiD_Scatterplots_Theta}
    \end{subfigure}
    \caption{\setstretch{0.5}(a) Resolutions for a distribution of electron energies incident in and out of the overlapping region in $\phi$. Each point represents the average resolution across 400 events. The 16+8 layer setup was corrected using a NN trained on electrons with energies between 20 and 300GeV and with angle ranges $\theta$ = 0$^{\circ}$-45$^{\circ}$, $\phi$ = 0$^{\circ}$-360$^{\circ}$. The left graph shows a non-overlapping part of the ECal ($\phi$ =  17$^{\circ}$-23$^{\circ}$), while the right graph shows most of the overlapping region ($\phi$ =  6$^{\circ}$-11$^{\circ}$). The corrected 16+8 design performed comparably to the 20+10 design in and out of the overlapping region across all energies. The corrected 16+8 design did not perform better than the uncorrected 16+8 design in the overlapping region due to finer sampling there. (b) Uncorrected percent deviations for ideal energy deposits for all $\theta$ values plotted as a function of $\phi$. There is significantly decreased deviation from the ideal deposit in the overlapping region due to finer sampling there. (c) Distorted energy profile caused by not separating energy deposits by module. (d) Corrected and uncorrected percent deviations from ideal deposit illustrating consistent NN correction across all electron energies.}
\end{figure}

We evaluated the NN correction on a distribution of both energies (20-300GeV) and angles (incidence angle $\theta$ = 0$^{\circ}$-45$^{\circ}$, azimuthal $\phi$ = 0$^{\circ}$-360$^{\circ}$) using a modified version of the full SiD simulation. Phi values were altered ($\phi \leftarrow \phi$ mod $30^{\circ}$) to move all events into the range $\phi$ = 0$^{\circ}$-30$^{\circ}$. Table \ref{tab:discreteEnergy_FullSiD_GroupedModules} shows the NN specifications used. The NN was trained on the energy deposits from each of the 32 silicon layers (including simulated layers), 24 layers' hit multiplicities (not including simulated layers), the total energy in 32 layers, $\theta$, and $\phi$.

The SiD model was modified to remove all systems except the ECal in order to ensure that we know the true electron energy and incidence angle (no bremsstrahlung or curving in magnetic field). The ECal was also extended to 60 layers to evaluate the effectiveness of the correction against a theoretically ideal ECal (see figure \ref{fig:ECalOverlap}).

In calculating which deposit was in which layer for the purposes of determining total energy deposit per layer, which module the deposit was in was not considered. This caused distorted shower profiles in the overlapping regions where two modules meet and where a single shower spans two modules (see figure \ref{fig:FullSiD_DistortedProfile}). The preliminary NN correction did not attempt to alter these distorted energy profiles. Instead of using the best-fit formula $f(t)=Ct^{\alpha}e^{-\beta t}$ to approximate deposits in odd-numbered ECal layers (which would not be accurate for these distorted profiles), it approximated each odd-numbered layer deposit as the average of neighboring even-numbered deposits. Figure \ref{fig:FullSiD_Scatterplots_Theta} shows the performance of this preliminary NN correction. Despite the altered shower profiles, the NN was able to correct most events to a percent deviation from ideal deposit of less than 5\%, especially for high electron energies.

Figure \ref{fig:FullSiD_Resolutions} shows how the corrected and uncorrected resolutions differed between the overlapping and non-overlapping regions. In the non-overlapping regions, the NN correction provided substantial resolution improvement, especially at high energies, providing corrected 16+8 resolutions comparable to those of the uncorrected 20+10 design. In the overlapping region, the NN provided no improvement in resolution. However, due to the fact that the sampling layers are closer together in the overlapping region, the uncorrected 16+8 resolution is already comparable to the 20+10 uncorrected resolution.

%For plots showing corrected and uncorrected scaled total energy deposits, from which the resolutions in figure \ref{fig:FullSiD_Resolutions} were calculated, see appendix G.

\subsection{Neural network correction - separation of modules in modified full SiD model}

\begin{table}[]
    \centering
    \begin{tabular}{|c|c|}
        \hline
        Network type & Fully connected \\
        \hline
        Hidden layers & 1 (163 nodes) \\
        \hline
        Free parameters & 26896 \\
        \hline
        Activation function & ReLu \\
        \hline
        Optimizer & Adam \\
        \hline
        Loss & Mean squared error \\
        \hline
        Libraries & Keras with TensorFlow backend \\
        \hline
    \end{tabular}
    \caption{Neural network configuration used to correct distributions of electron energies in a modified SiD ECal design fired at random energies (20-300GeV) and angles ($\theta$ = 0$^{\circ}$-45$^{\circ}$, $\phi$ = 0$^{\circ}$-360$^{\circ}$) using a modified version of the full SiD simulation while separating the modules of events with deposits in multiple modules.}
    \label{tab:discreteEnergy_FullSiD_SeparatedModules}
\end{table}

\begin{figure}
    \centering
    \begin{subfigure}{\linewidth}
        \centering
        \hspace*{-0.5in}
        \includegraphics[scale=0.5]{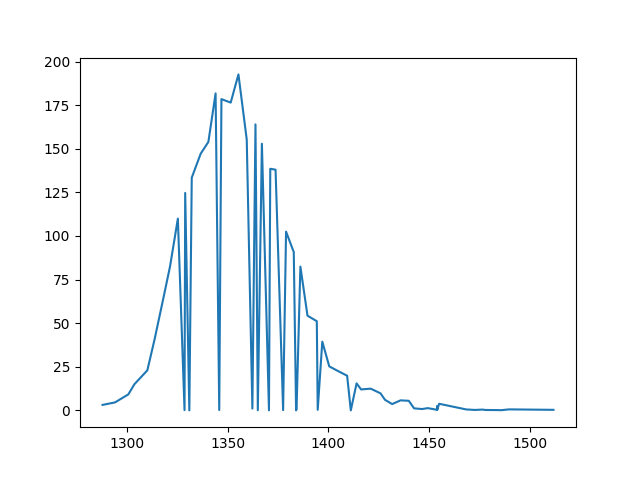}
        \hspace*{-0.8cm}
        \includegraphics[scale=0.5]{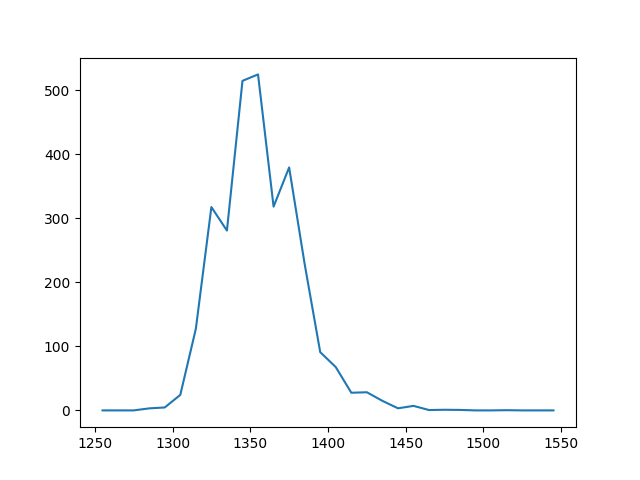}
        \caption{Energy deposits (MeV) for single electron event with $\phi = 17.38^{\circ}$ plotted against radius (mm). The left graph shows individual deposits, while the right graph shows deposits grouped in bins of size 10mm.}
        \label{fig:FullSiD_SeparatedModules_AvgAndNonAvg}
    \end{subfigure}
    \begin{subfigure}{\linewidth}
        \hspace*{-0.1in}
        \includegraphics[scale=0.5]{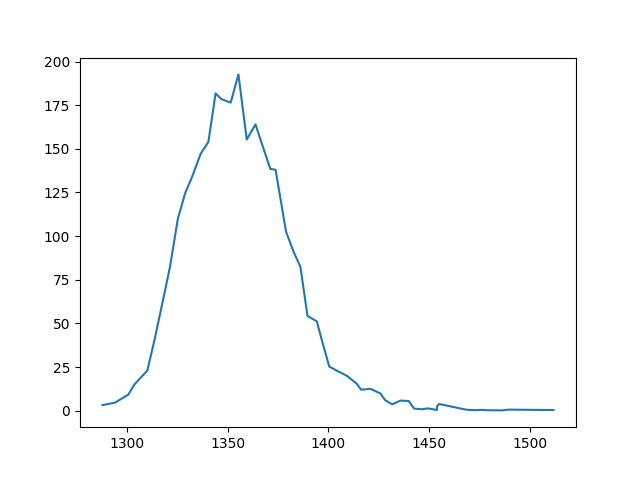}
        \hspace*{-0.5cm}
        \includegraphics[scale=0.5]{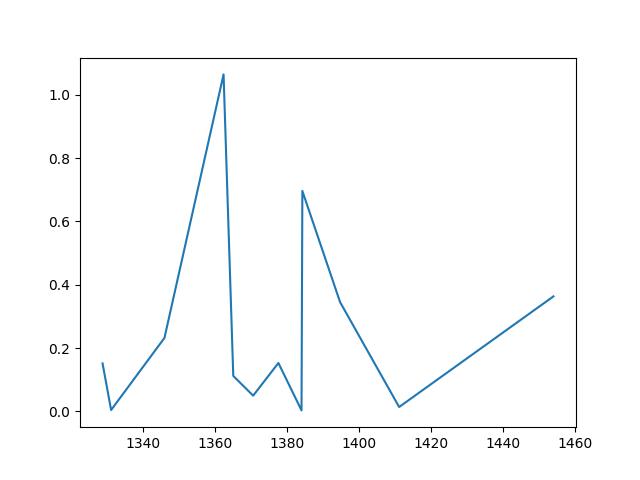}
        \caption{Energy deposits (MeV) for single electron event with $\phi = 17.38^{\circ}$ plotted against radius (mm) and separated by module. Most of the energy is deposited in a single module.}
        \label{fig:FullSiD_SeparatedModules_MostlyOneModule}
    \end{subfigure}
    \begin{subfigure}{\linewidth}
        \hspace*{-0.1in}
        \includegraphics[scale=0.5]{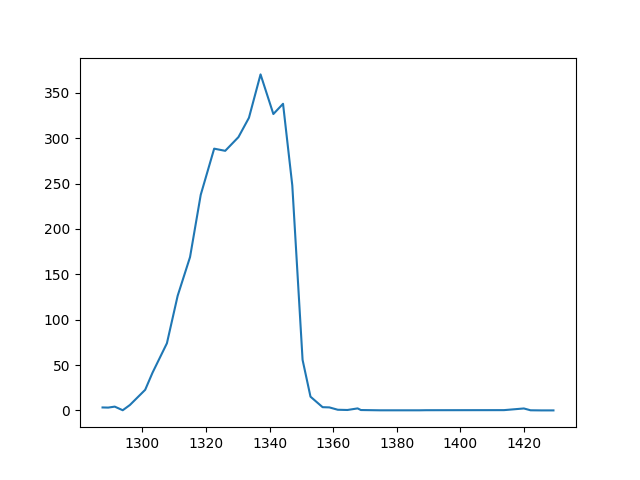}
        \hspace*{-0.5cm}
        \includegraphics[scale=0.5]{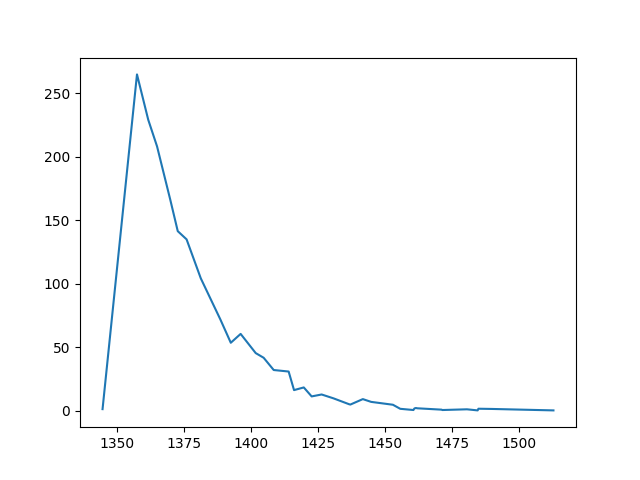}
        \caption{Energy deposits (MeV) for single electron event with $\phi = 9.27^{\circ}$ plotted against radius (mm) and separated by module. Highly significant energy deposits appear in both modules.}
        \label{fig:FullSiD_SeparatedModules_EvenlySplit}
    \end{subfigure}
    \caption{Energy profiles plotted in different ways. (a) Plotting all individual deposits leads to a messy energy profile; this can be fixed by binning but not without losing information about the deposits in each individual module. (b) and (c) Energy profiles separated by module. This removes the need for binning while letting us tell how much energy was deposited in each module.}
    \label{separate_modules}
\end{figure}

\begin{figure}
    \centering
    \begin{subfigure}{\linewidth}
        \hspace*{-0.5in}
        \includegraphics[scale=0.4]{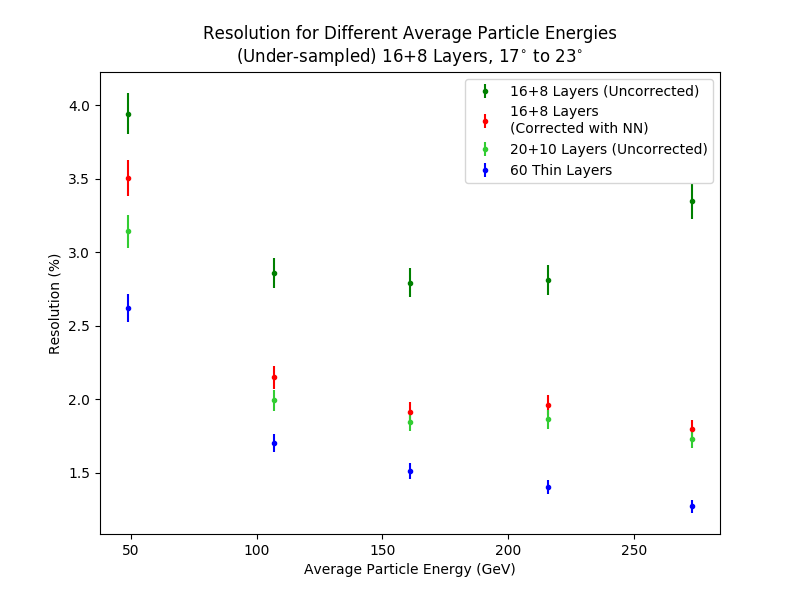}
        \hspace*{-0.8cm}
        \includegraphics[scale=0.4]{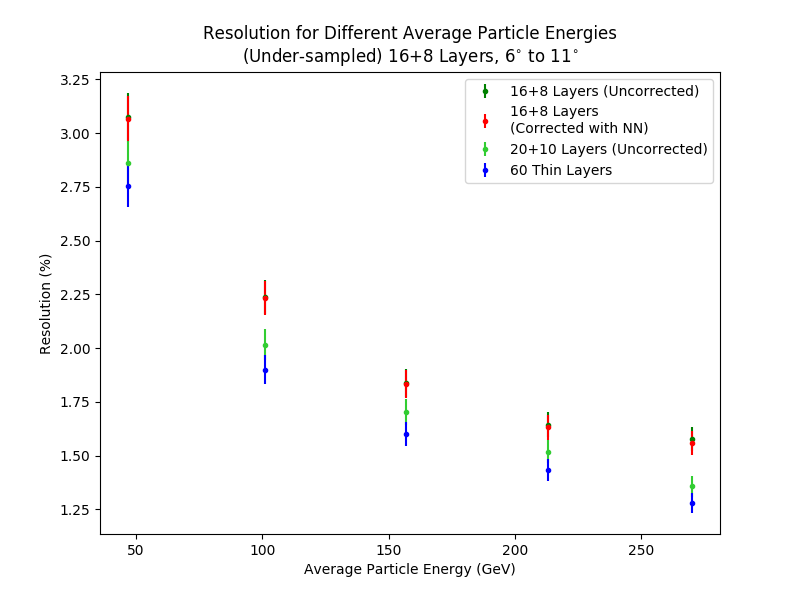}
        \caption{}
        \label{fig:FullSiD_Resolutions_SeparatedModules}
    \end{subfigure}
    \begin{subfigure}{\linewidth}
        \hspace*{-0.1in}
        \includegraphics[scale=0.45]{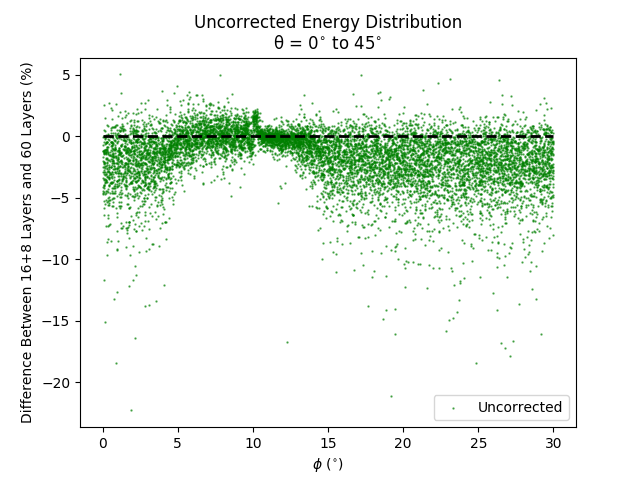}
        \hspace*{-0.5cm}
        \includegraphics[scale=0.45]{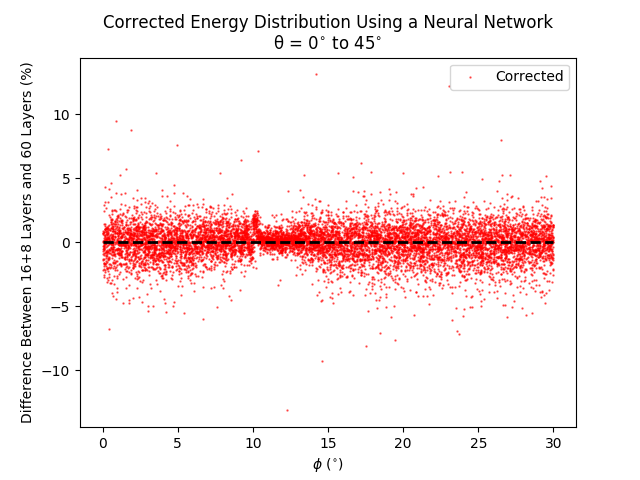}
        \caption{}
        \label{fig:FullSiD_Scatterplots_Phi_SeparatedModules}
    \end{subfigure}
    \begin{subfigure}{\linewidth}
        \hspace*{-0.1in}
        \includegraphics[scale=0.45]{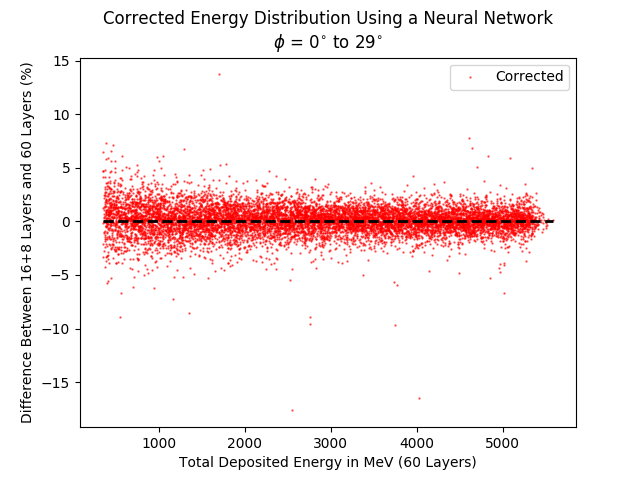}
        \hspace*{-0.5cm}
        \includegraphics[scale=0.45]{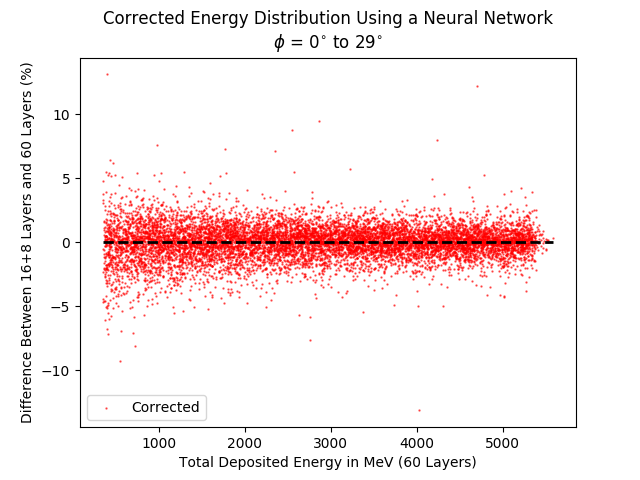}
        \caption{}
        
        \begin{tikzpicture}[remember picture,overlay]
          \node[anchor=south west,inner sep=0pt] at ($(current page.south west)+(6.5cm,11cm)$) {
       \textbf{Unseparated}
        };
     \end{tikzpicture}
     \begin{tikzpicture}[remember picture,overlay]
          \node[anchor=south west,inner sep=0pt] at ($(current page.south west)+(14cm,11cm)$) {
       \textbf{Separated}
        };
     \end{tikzpicture}
     \label{fig:FullSiD_Scatterplots_OldAndNew}
    \end{subfigure}
    \caption{\setstretch{0.5}(a) Resolutions for a distribution of electron energies incident in and out of the overlapping region in $\phi$. Each point represents the average resolution across 400 events. The 16+8 layer setup was corrected using a NN trained on electrons with energies between 20 and 300GeV and with angle ranges $\theta$ = 0$^{\circ}$-45$^{\circ}$, $\phi$ = 0$^{\circ}$-360$^{\circ}$. Energy profiles were split by module to perform a NN analysis. The left graph shows a non-overlapping part of the ECal ($\phi$ =  17$^{\circ}$-23$^{\circ}$), while the right graph shows most of the overlapping region ($\phi$ =  6$^{\circ}$-11$^{\circ}$). Performance of the NN correction was comparable to that in figure \ref{fig:FullSiD_Resolutions}. (b) Corrected and uncorrected percent deviations from ideal deposit as a function of $\phi$. The NN consistently corrected the percent deviation for all phi values. (c) Corrected percent deviations from ideal deposit with (right) and without (left) separating modules. The separate modules performed roughly equivalently to the unseparated modules.}
    \label{separated_modules_NN}
\end{figure}

To improve the ease of analyzing shower profiles and to attempt to improve NN performance, we calculated energy deposits per layer for each module separately (almost all events deposited energy in exactly two different modules). To effectively represent the deposits from each module, several methods were considered. Figure \ref{fig:FullSiD_SeparatedModules_AvgAndNonAvg} shows how simply plotting all individual deposits as a function of radius can lead to messy energy profiles when most of the energy was deposited in a single layer, as well as how this can be solved by grouping deposits within bins (size 10mm). However, binning prevents effective analysis of how much energy was deposited in each layer and so was not used. Figures \ref{fig:FullSiD_SeparatedModules_MostlyOneModule} and \ref{fig:FullSiD_SeparatedModules_EvenlySplit} show energy profiles where the deposits were separated out by module before calculating energy deposit per layer. This allows for more effective shower analysis when substantial energy is deposited in each module (as in figure \ref{fig:FullSiD_SeparatedModules_EvenlySplit}) while preventing small deposits from influencing overall analysis (as in figure \ref{fig:FullSiD_SeparatedModules_MostlyOneModule}). We ultimately used a separating method for NN analysis in which deposits from each module were fed separately into a NN with an increased number of nodes.

The NN was trained on the energy deposits from each of the 64 silicon layers (including simulated layers), 48 layers' hit multiplicities (no simulated layers), the distance from the center of the detector for 48 layers (no simulated layers), the total energy in 32 layers, $\theta$, and $\phi$. The specifications for the NN used can be found in table \ref{tab:discreteEnergy_FullSiD_SeparatedModules}.

Similar to the preliminary full SiD NN correction, the separate-module NN correction approximated each odd-numbered layer deposit as the average of neighboring even-numbered deposits (averaging method) instead of using the best-fit formula $f(t)=Ct^{\alpha}e^{-\beta t}$ to approximate deposits in odd-numbered ECal layers. This is because, despite the cleaner shower profiles, small-deposit layers would not properly follow the pattern implied by the fit. Figure \ref{fig:FullSiD_Resolutions_SeparatedModules} shows corrected and uncorrected resolution values in the overlapping (right) and non-overlapping (left) regions. The resolution improvement when calculating leakage treating each module separately is comparable to when calculating without separating modules and shows the same trend of strong improvement in the non-overlapping region and no improvement (and little need for improvement) in the overlapping region (see figure \ref{fig:FullSiD_Resolutions}). Figure \ref{fig:FullSiD_Scatterplots_OldAndNew} similarly shows low values for percent deviations from the ideal energy deposit when treating modules separately, but shows no significant difference from when modules are not treated separately.

It should be possible to further improve the NN correction through a more sophisticated method of simulating odd-numbered layers when treating modules separately. We have not done so here, but we do note one complications with doing so: the theory-informed best fit methodology $f(t)=Ct^{\alpha}e^{-\beta t}$ cannot always be used when not all energy deposits occur in the same module. Thus, a cutoff should be developed for what fraction of the energy needs to be deposited in a single module before $f(t)$ becomes a sufficiently accurate best fit. Additionally, as shown in figure \ref{fig:FullSiD_Resolutions_SeparatedModules}, the preliminary averaging method yields resolutions comparable to the 20+10 uncorrected resolutions, so large further improvement through a sophisticated best fit is unlikely.

\section{Conclusion}
Electron energy deposits in the SiD ECal follow patterns which can be used to estimate the amount of energy which leaks out of the calorimeter. We developed first-order and neural-network (NN) methodologies for improving the accuracy of ECal electron energy measurements by analyzing shower patterns to predict energy leakage. Emphasis was placed on evaluating the possibility of reducing the size of the ECal while maintaining low resolution using leakage correction.

Leakage correction was first performed on a Geant4 simple stack \cite{allison_geant4_2006,allison_recent_2016}. First-order corrections on the reduced ECal (16+8 design) using the shower max layer and the fraction of energy deposited near the back of the calorimeter accurately corrected the mean deposit for all energies from 20GeV to 250GeV. However, the resolution was significantly worse than the nominal (20+10 design) resolution for high energies (above 200GeV), see figure \ref{fig:discrete_discrete}. A preliminary neural network correction improved on this resolution so that the 16+8 corrected and 20+10 uncorrected resolutions were comparable (figures \ref{fig:50and100GeV_discreteEnergy_discreteAngle_distTraining} and \ref{fig:Resolutions_discreteEnergy_discreteAngle_distTraining}). The NN also reduced the bias in the correction by training on a continuous flat distribution of electron energies instead of discrete electron energies.

We further trained and tested a neural network on electrons incident at random angles (0$^{\circ}$-50$^{\circ}$) and random energies (20GeV-300GeV) to better emulate realistic experimental conditions. At shallow angles, the NN greatly improved the resolution and mean energy measurement. At steep angles, the resolution did not improve and the mean was corrected only slightly, but reduced leakage at steep angles made correction less necessary (figure \ref{fig:distEnergy_distAngle}).

We used an altered version of the DD4hep SiD detector \cite{frank_dd4hep:_2014} description to test the NN correction in the full SiD ECal model. The overlapping region in $\phi$ caused the shower to occur in multiple modules, distorting the energy profile (figures \ref{fig:FullSiD_DistortedProfile} and \ref{fig:ECalOverlap}). However, the NN correction still performed well due to the finer sampling in the overlapping region reducing the need for leakage correction (figure \ref{fig:FullSiD_Resolutions}). To see how showers were split across multiple modules, deposits in each module were separated. This allowed for better analysis of shower development but did not alter the NN performance (figures \ref{separate_modules} and \ref{separated_modules_NN}).

Leakage correction using a neural network thus allows for the corrected reduced ECal design to function comparably to the uncorrected nominal ECal design for a wide range of electron energies (20GeV-300GeV) and angles ($\theta$ = 0$^{\circ}$-45$^{\circ}$, $\phi$ = 0$^{\circ}$-360$^{\circ}$). Thus, the nominal performance can be maintained with 24 silicon, 16 thin tungsten and 8 thick tungsten layers, instead of the nominal 30 silicon, 20 thin tungsten, and 10 thick tungsten layers. Based on the ILC TDR \cite{behnke_international_2013}, this would cause about a 21\% reduction in materials cost.

\section{Acknowledgements}
Special thanks the SiD Detector Optimization team for their strong feedback and guidance in this research. Acknowledgement is given to the University of Oregon for providing the necessary computing resources, as well as to the US DOE for their funding of SiD research. Acknowledgement is given to A. Steinhebel for the graphic in figure \ref{fig:ECalOverlap}.

\bibliographystyle{aipnum4-1}

%%% BIBLIOGRAPHY CONTENTS %%%


\end{document}